\documentclass[useAMS,usenatbib]{mn2e}
\usepackage[british]{babel}

\usepackage{graphicx}

\usepackage{amsmath}
\usepackage{mathabx}
\usepackage{mathtools}

\title[Angular momentum of galactic coronae]{The angular momentum of cosmological coronae and the inside-out growth of spiral galaxies}

\author[Pezzulli et al.]{Gabriele Pezzulli$^{1, 2, 3}$\thanks{E-mail:gabriele.pezzulli@phys.ethz.ch}, Filippo Fraternali$^{3, 4}$ and James Binney$^2$\\
$^{1}$ Institute for Astronomy, ETH Zurich, Wolfgang-Pauli-Strasse 27, CH-8093 Zurich, Switzerland \\
$^{2}$ Rudolf Peierls Centre for Theoretical Physics, University of Oxford, 1 Keble Road, Oxford OX1 3NP, UK \\
$^{3}$ Department of Physics and Astronomy, University of Bologna, Viale Berti Pichat 6/2, I-40127 Bologna, Italy \\
$^{4}$ Kapteyn Astronomical Institute, University of Groningen, Postbus 800, NL-9700 AV, Groningen, The Netherlands \\}

\begin{document}

\date{Accepted 2017 January 05. Received 2017 January 05; in original form 2016 July 03}

\pagerange{\pageref{firstpage}--\pageref{lastpage}} \pubyear{2017}

\maketitle

\label{firstpage}

\begin{abstract}
\noindent
Massive and diffuse haloes of hot gas (\emph{coronae}) are important intermediaries between cosmology and galaxy evolution, storing mass and angular momentum acquired from the cosmic web until eventual accretion on to star forming discs. We introduce a method to reconstruct the rotation of a galactic corona, based on its angular momentum distribution (AMD). This allows us to investigate in what conditions the angular momentum acquired from tidal torques can be transferred to star forming discs and explain observed galaxy-scale processes, such as inside-out growth and the build-up of abundance gradients. We find that a simple model of an isothermal corona with a temperature slightly smaller than virial and a cosmologically motivated AMD is in good agreement with galaxy evolution requirements, supporting hot-mode accretion as a viable driver for the evolution of spiral galaxies in a cosmological context. We predict moderately sub-centrifugal rotation close to the disc and slow rotation close to the virial radius. Motivated by the observation that the Milky Way has a relatively hot corona $(T \simeq 2 \times 10^6 \; \textrm{K})$, we also explore models with a temperature larger than virial. To be able to drive inside-out growth, these models must be significantly affected by feedback, either mechanical (ejection of low angular momentum material) or thermal (heating of the central regions). However, the agreement with galaxy evolution constraints becomes, in these cases, only marginal, suggesting that our first and simpler model may apply to a larger fraction of galaxy evolution history.
\end{abstract}


\section{Introduction}\label{sec::Introduction}

Observational clues for the existence of large haloes of hot ($T \sim 10^6 \; \textrm{K}$) diffuse ($n \sim 10^{-4} \; \textrm{cm}^{-3}$) gas (\emph{coronae}) around massive (Milky-Way-like) spiral galaxies date back to \cite{Spitzer1956}, although these structures have been unambiguously detected, in X-ray emission, only recently (e.g.\ \citealt{AndersonBregman2011}; \citealt{Dai+12}; \citealt{Bogdan+13a}; \citealt{Bogdan+13b}; \citealt{MillerBregman2015}). It seems plausible that galactic coronae extend to radii comparable to the virial radius and contain a mass equal to or larger than that of the embedded galaxies (e.g.\ \citealt{FP06}; \citealt{Gatto+13}; \citealt{Anderson+16}). Structures with these properties had also been predicted by early galaxy formation theories (\citealt{RO77}; \citealt{WR78}), although the initial picture has since then been shown to depend quite strongly on the details of radiative cooling (\citealt{Binney77}; \citealt{BirnboimDekel2003}; \citealt{DB06}) and feedback (e.g.\ \citealt{Binney2004}; \citealt{Cantalupo2010}). One clear point is that, in order to exist at the present epoch, galactic coronae must cool inefficiently (either because the cooling times are large, or because cooling is somehow compensated or inhibited), on time-scales comparable to the Hubble time. As a consequence, it is plausible that they condense on to the galaxies at their centres only very gradually, sustaining, for cosmological time-scales, the moderate but persistent levels of star formation that are commonly inferred for the discs of spiral galaxies (e.g. \citealt{FT12}; \citealt{Pacifici+16}). The reason why coronal condensation is much more effective for spirals than for ellipticals is still unclear and could be related either to cooling-suppressing mechanisms being especially powerful in early-type galaxies (e.g.\ \citealt{McNamara+07}; \citealt{Ciotti+10}; \citealt{Gaspari+12}), or to some form of positive star formation feedback active in spirals (e.g.\ \citealt{Marinacci+10}; \citealt{Hobbs+15}; \citealt{Armillotta+16}), or to both effects.

If galaxy discs acquire a significant part of their mass from coronae, then coronae must contain much angular momentum. In fact, accreting mass without accreting angular momentum would lower the average specific angular momentum of a disc, forcing it to shrink with time. This is at variance with discs of spiral galaxies actually \emph{increasing} in size while growing in mass (\emph{inside-out growth}), as supported by a wide variety of observations (e.g. \citealt{Barden+05}; \citealt{Gogarten+10}; \citealt{MMIII}; \citealt{CALIFA}), by chemical evolution models (e.g.\ \citealt{BP99}; \citealt{Chiappini}; \citealt{MD05}) and by cosmological simulations (e.g.\ \citealt{Pilkington+12}; \citealt{AWN2014}). Moreover, independent observations indicate inside-out growth to be a continuous process, stil active today for most star-forming galaxies (e.g.\  \citealt{Williams+09}; \citealt{Wang+11}; \citealt{Pezzulli+15}; \citealt{Ibarra-Medel+16}), which suggests that these galaxies are in continuous need for fresh supply of high-angular momentum gas. As all extended structures, coronae will have acquired angular momentum from cosmological tidal torques (e.g.\ \citealt{Peebles}; \citealt{BarnesEfstathiou1987}). Taking into account that dark matter and baryons should have experienced similar tidal torques until the moment of their incorporation on galactic haloes (e.g.\ \citealt{FallEfstathiou1980}), the \emph{global} specific angular momentum of a typical galactic corona should be similar to that of the dark matter (or slightly larger, see e.g.\ \citealt{Teklu+15} and references therein), which has been estimated to be somewhat larger than that of the disc (\citealt{RF12}; \citealt{DvdB12}). There is therefore, in principle, enough angular momentum, in the hot gaseous haloes, to sustain the ongoing inside-out growth of discs. Whether this angular momentum is \emph{distributed}, within these huge structures, in such a way to be actually available to sustain disc growth is, however, far from obvious. Addressing this problem, with relatively simple calculations, is the main goal of this paper.

\subsection{Preliminary considerations}\label{subsec::preliminary}
For inside-out growth to be continuously fed by coronal accretion, the disc must be, at any time, in contact with coronal material with a specific angular momentum $l$ that is larger than the \emph{average} specific angular momentum of the disc $l_\textrm{disc}$. Dynamical stability bounds the specific angular momentum $l$ of the corona to increase with cylindrical radius $R$ and therefore the relevant material is stored outside some \emph{crossing radius} $R_\textrm{cross}$ defined by
\begin{equation}\label{Rcrossdef}
l(R_\textrm{cross}) = l_\textrm{disc} \;.
\end{equation}
If $R_\textrm{cross}$ is small (comparable to the disc size), coronal gas with sufficiently large angular momentum is in contact with the galaxy and, if the physical conditions for condensation are met, the accretion of this gas can sustain the inside-out growth of the disc. Conversely, if $R_\textrm{cross}$ is large (comparable to the virial radius), the disc is only in contact with gas with $l < l_\textrm{disc}$, the accretion of which would cause the disc to shrink.

A small $R_\textrm{cross}$ requires a relatively large rotation velocity of the corona (comparable to that of the disc), at least in the inner regions. Observations of coronal rotation are very challenging, mainly because of the poor spectral resolution of the current X-ray instrumentation (\citealt{MHB16}). Nonetheless, recent estimates (\citealt{HB16}) indicate that the \emph{inner} corona of the Milky Way rotates at $V_\textrm{cor} \simeq 180 \; \textrm{km} \; \textrm{s}^{-1} $: this is smaller than, but comparable to the rotation velocity of the Galactic disc ($V_\textrm{disc} \simeq 240 \; \textrm{km} \; \textrm{s}^{-1}$, \citealt{Schoenrich2012}). The fact that $V_\textrm{cor} < V_\textrm{disc}$ is perfectly consistent with basic hydrodynamic considerations: if the gas is hot enough, pressure gradients can provide significant support against gravity, implying lesser need of centrifugal support and slower rotation than that of the cold gas in the disc. On the other hand, the fact that the lag is \emph{small} ($V_\textrm{cor} \lesssim V_\textrm{disc}$, rather than $V_\textrm{cor} \ll V_\textrm{disc}$) implies that, at least in the inner regions, the corona is still significantly supported by rotation and has a dynamically significant angular momentum. A small, though non-vanishing, difference between $V_\textrm{cor}$ and $V_\textrm{disc}$ is also independently suggested by chemical evolution arguments. It has long been known that any difference between the rotation velocity of the disc and that of the gas accreting on it inevitably induces a systematic radial gas flow within the disc, with a strong and testable impact on chemical evolution (\citealt{MV81}; \citealt{LF85}; \citealt{PT89}). Recent studies of this effect (\citealt{BS12}; \citealt{PF16}) have predicted that, to match the chemical properties of our Galaxy, the hot gas around the Milky Way must rotate at $\sim 70 - 80 \%$ of the rotation velocity of the disc, in very good agreement with the later observational findings of \cite{HB16}.

A scenario in which $V_\textrm{cor} \lesssim V_\textrm{disc}$, at least near to the disc, is therefore supported by independent arguments. On the other hand, a condition of \emph{near corotation} cannot be extrapolated to the whole structure of the corona; because the corona is far more extended than the disc, $V_\textrm{cor} \simeq V_\textrm{disc}$ everywhere would imply a global angular momentum of the corona far in excess of what is allowed by tidal torque theory. The simple picture therefore arises that, to be compatible with both galaxy evolution and cosmology, the corona should rotate relatively fast (mildly sub-centrifugally) in the inner regions (on scales comparable to size of the disc, say $\sim 10 \; \textrm{kpc}$) and much slower in the outer ones (on scales comparable to the virial radius, say $\sim 100 \; \textrm{kpc}$). This is a prediction, which will be confirmed or refuted by future observations of the kinematics of coronae around the Milky Way and external galaxies (see \citealt{MHB16} on future observational prospects in this direction).

While waiting for observational advances, we can make a step forward and try to understand whether, in addition to the collection of arguments given above, there is also a coherent physical picture which can \emph{explain} why galactic coronae should rotate in that particular way. One way to address this question is to analyse the output of cosmological hydrodynamical simulations. \cite{Teklu+15} find that the specific angular momentum of the hot gas rises steeply with radius in the inner regions and more slowly in the outer parts, as expected for a declining rotation velocity profile. More studies of this kind would be useful to understand the robustness of these results with respect to the adopted numerical scheme and on prescriptions of sub-grid physics and in particular feedback. In this paper, we try to address the same problem from an analytic point of view. 

Our starting point is the \emph{angular momentum distribution} (AMD) of the corona, defined as the amount of mass $M$ per unit specific angular momentum $l$:
\begin{equation}\label{defpsi}
\psi \vcentcolon = \frac{dM}{dl} \; .
\end{equation}
The zeroth and first order moments of $\psi$ coincide with the total mass and average specific angular momentum, respectively:
\begin{equation}\label{simplechoice}
\begin{split}
M_1 & = \int_0^{+\infty} \psi(l) dl \\
l_1 & = \frac{1}{M_1} \int_0^{+\infty}l\psi(l)dl \; ,
\end{split}
\end{equation}
while the \emph{shape} of the function contains a much larger (in principle, infinitely larger) amount of additional independent information, which can be used to study the details of rotating structures. Classical studies (e.g.\ \citealt{Mestel1963}; \citealt{FallEfstathiou1980}) have made ample use of this idea to understand the problem of galaxy formation. More recently, the AMD has been a key in clarifying the local nature of the angular momentum problem in the formation of galaxy discs (e.g.\ \citealt{vdB+01}; \citealt{MallerDekel2002}). Here we use the AMD, together with the equations of rotating hydrodynamic equilibrium, to break the local degeneracy between pressure support and rotational support: this allows us to reconstruct what the density and rotation velocity should be, as a function of radius, for a given AMD. We initially assume that the gas has a primordial AMD and is in isothermal equilibrium in a logarithmic potential, then we investigate the consequences of progressively relaxing these assumptions. We assume throughout barotropic equilibrium (i.e.\ that pressure and density are stratified on the same surfaces). The most general case of baroclinic (i.e.\ non barotropic) equilibrium is left for a future investigation.

\subsection{Overview}
The rest of the paper is structured as follows. In Section \ref{sec::Basics} we introduce our formalism. In Section \ref{sec::inversion} we explain our method to reconstruct a configuration of rotating equilibrium given its AMD. In Section \ref{sec::CosmoAMD}, we apply our method to a cosmologically motivated AMD and compare model predictions with the requirements of coronal-accretion-driven inside-out growth. In Sections \ref{sec::SuperVirialModels}, \ref{sec::WindModels} and \ref{sec::PoliModels} we explore how our models respond to some observationally and theoretically motivated changes on the coronal temperature and AMD and to deviations from isothermality. In Section \ref{sec::Potential} we demonstrate the robustness of our approach with respect to the choice of the gravitational potential. Section \ref{sec::Discussion} further discusses the consequences and the limits of our investigation. Our findings and conclusions are summarized in Section \ref{sec::Summary}.

\section{Basic equations and models}\label{sec::Basics}

\subsection{Starting equations}\label{sec::starteq}
Let the corona around a spiral galaxy be modelled, in first approximation, as isothermal gas in rotating equilibrium in a logarithmic potential:
\begin{equation}\label{logpot}
\Phi(R, z) = V_d^2 \ln \left( \frac{\sqrt{R^2+z^2}}{\tilde{r}} \right) \; ,
\end{equation}
where $(R, z)$ are cylindrical coordinates, $\tilde{r}$ is an arbitrary radius and $V_d$ is the (constant) rotation velocity of a disc in centrifugal equilibrium in the potential \eqref{logpot}. We will relax the isothermal hypothesis in Section \ref{sec::PoliModels} and discuss the other limits of our assumptions in Sections \ref{sec::Potential} and \ref{sec::Discussion}.

If the density in the equatorial plane at some radius $R$ is equal to $\rho_0(R)$, then, from vertical hydrostatic equilibrium, we have
\begin{equation}\label{verticaleq}
\rho(R, z) = \rho_0(R) e^{- \frac{\Phi(R, z)-\Phi(R, 0)}{c_s^2} } = \rho_0(R) \left( 1 + \left( \frac{z}{R} \right)^2 \right)^{- \frac{a}{2}} \; ,
\end{equation}
where $c_s^2$ is the isothermal `sound speed' squared (defined as the constant ratio between pressure and density) and the constant
\begin{equation}\label{adef}
a \vcentcolon = \left(\frac{V_d}{c_s}\right)^2 = \frac{3 T_\textrm{vir}}{T}
\end{equation}
is inversely proportional to the temperature $T$ of the corona. For the \emph{virial temperature} $T_\textrm{vir}$ we adopt the definition
\begin{equation}\label{Tvirdef}
T_\textrm{vir} \vcentcolon = \frac{\mu m_p}{3 k_B}  V_d^2 = 1.37 \left( \frac{V_d}{240 \; \textrm{km} \; \textrm{s}^{-1}} \right)^2 10^6 \; \textrm{K} \; ,
\end{equation}
where $k_B$ is the Boltzmann constant and $\mu = 0.59$ is the average mass per particle in units of the mass of the proton $m_p$, as appropriate for a completely ionized plasma with helium mass abundance $Y = 0.25$. Note that $T_\textrm{vir}$ is the temperature at which a non-rotating isothermal gas in equilibrium in the logarithmic potential \eqref{logpot} has a kinetic energy equal to the one prescribed by the virial theorem, in the absence of additional forces \footnote{This is immediately seen if the virial theorem is stated in the form $\overline{K} = \overline{ x \cdot \nabla \Phi}/2$, where bars indicate averages, the dot indicates a scalar product, $K$ is kinetic energy, $\Phi$ the potential and $x$ the position vector (see e.g.\ equation 3-27 of \citealt{Goldstein}). In the absence of rotation, the average kinetic energy per particle equals that of thermal motion, which is $3 k_B T/2$. If the gas rotates, the kinetic energy of thermal motion is only a part of the total kinetic energy and therefore the temperature must be smaller than the virial value, unless the system is subject to additional forces.\label{foot::virial}}.

The projected surface density at radius $R$ is
\begin{equation}\label{Sigma}
\Sigma(R) = 2 \int_0^{+\infty} \rho(R, z) dz = 2\chi R\rho_0(R) \; ,
\end{equation}
where the dimensionless constant
\begin{equation}\label{defk}
\chi = \int_0^{+\infty} (1 + x^2)^{-a/2} dx
\end{equation}
is finite for $a>1$, or $T < 3 T_\textrm{vir}$, which sets a first upper limit to the temperature of the corona for an isothermal solution to exist in a logarithmic potential. We will see, however, that a more stringent limit comes from requiring a finite total mass.

The isothermal assumption implies that the rotation velocity $V$ of the corona is a function of the (cylindrical) radius $R$ only (the \emph{Poincar\'e-Wavre theorem}, e.g.\ \citealt{Tassoul2000}). Such a function also determines the local shape of the equatorial density distribution $\rho_0$ by means of the radial equation of hydrodynamic equilibrium in the equatorial plane, which can be written:
\begin{equation}\label{Euler}
\frac{d \ln \rho_0}{d\ln R} (R) = a (w^2(R) -1) \; ,
\end{equation}
where
\begin{equation}\label{wdef}
w(R) \vcentcolon = \frac{V(R)}{V_d} = \frac{l(R)}{RV_d}
\end{equation}
is the ratio of the rotation velocity $V$ of the corona to the rotation velocity $V_d$ of the disc, while
\begin{equation}
l(R) = RV(R)
\end{equation}
is the specific angular momentum of the corona at radius $R$. Note in particular from equation \eqref{Euler} that the corona rotates slower than the disc whenever its equatorial density is a decreasing function of radius. 

From equation \eqref{Sigma}, the mass enclosed inside a cylindrical surface of radius $R$ obeys
\begin{equation}\label{masseq}
\frac{dM}{dR} (R)= 2\pi R \Sigma(R) = 4 \pi \chi R^2 \rho_0(R) \; .
\end{equation}
Taking equations \eqref{Euler} and \eqref{masseq} into account, mass convergence at large radii requires
\begin{equation}
-3 > \lim_{\substack{R \to \infty}} \frac{d \ln \rho_0}{d\ln R} (R) \geq -a \; ,
\end{equation}
which implies $a > 3$, or $T < T_\textrm{vir}$. This again indicates the virial value as an upper limit to the coronal temperature, as already derived from the virial theorem (see footnote \ref{foot::virial}). Note, however, that equilibrium models with a super-virial temperature ($1< T/T_\textrm{vir} < 3$) can also be conceived, but they require truncation at finite radius and pressure confinement by an external medium. The latter obviously breaks the isolation of the system, with the consequence that the virial theorem does not apply.

Mass convergence at small radii, on the other hand, sets a lower limit to the central rotation velocity of the corona: 
\begin{equation}\label{V0_LowerLimit}
w(0) \geq \sqrt{1 - \frac{3}{a}} = \sqrt{1 - \frac{T}{T_\textrm{vir}}} \; ,
\end{equation}
which is well-defined in self-consistent sub-virial models, where $a > 3$ (we will discuss the super-virial case in Sections \ref{sec::HotBranch} and \ref{sec::Remark}). Note that in practice the `central' velocity $w(0)$ above is better understood as an estimate of the inner rise of the rotation of the corona, due to deviations of the potential from a logarithmic shape at small radii (see e.g.\ \citealt{Marinacci+11} and Section \ref{sec::Potential} below).

Note finally that the constant $\chi$, defined in equation \eqref{defk}, decreases monotonically from 1 to 0 as $a$ increases from 3 to $+ \infty$ (that is, as the temperature decreases from $T_\textrm{vir}$ to 0). Therefore, from equation \eqref{masseq}, the mass within the cylindrical radius $R$ never exceeds that contained within an equal \emph{spherical} radius by an ideal spherical distribution with radial profile $\rho_0$ (the equality holding for the virial temperature). The opposite happens for super-virial solutions ($1 < a < 3 $), where $\chi > 1$.

\subsection{Physical scales}\label{sec::dimensionless}
The equations in Section \ref{sec::starteq} can be put in dimensionless form if two physical scales are introduced alongside the velocity $V_\textrm{d}$ (or, equivalently, the virial temperature $T_\textrm{vir}$), associated with the gravitational potential \eqref{logpot}: the total mass $M_1$ and the average specific angular momentum $l_1$ of the corona, which are related to the AMD \eqref{defpsi} by equation \eqref{simplechoice}. We define, as a consequence, some scales for length and density:
\begin{equation}\label{scalings}
R_1 \vcentcolon= \frac{l_1}{V_d} \quad \rho_1 \vcentcolon= \frac{M_1}{4 \pi \chi R_1^3} \; ,
\end{equation}
and the dimensionless variables
\begin{equation}\label{dimensionlessvariables}
\lambda \vcentcolon= \frac{l}{l_1} \quad s \vcentcolon= \frac{R}{R_1} \quad y \vcentcolon= \frac{\rho_0}{\rho_1} \; .
\end{equation}
Note that the dimensionless rotation velocity $w$, defined in equation \eqref{wdef}, is related to the variables above by the equation
\begin{equation}\label{defw2}
w(s) = \frac{\lambda(s)}{s}
\end{equation}
and that the denominator in the definition of $\rho_1$ is chosen such that the mass contained within the dimensionless radius $s$ is
\begin{equation}
M(s) = M_1 \int_0^s \hat{s}^2 y(\hat{s}) d\hat{s} \; .
\end{equation}
Note that $R_1$ is the radius where the specific angular momentum of the disc equals the average specific angular momentum of the corona $l_1$.

As discussed in Section \ref{sec::Introduction}, $l_1$ is expected to be a few times the specific angular momentum of the disc $l_d$:
\begin{equation}\label{lineq}
l_1 \gtrsim l_d
\end{equation}
If the mass surface density of the disc is approximated by an exponential with scale-length $R_\textrm{d}$, then
\begin{equation}\label{expdiscldisc}
l_\textrm{d} = 2 V_\textrm{d} R_\textrm{d}
\end{equation}
and therefore, from equations \eqref{scalings}, \eqref{lineq} and \eqref{expdiscldisc}, the ratio
\begin{equation}\label{R1estimate}
\frac{R_1}{R_d} = 2\frac{l_1}{l_d}
\end{equation}
is also expected to be of the order of a few. The scaling radius $R_1$ is therefore comparable to the size of the disc. Since the corona is much larger than this, any sensible model must extend to some dimensionless radius $s_\textrm{max}$ significantly larger than unity.

For our fiducial Milky-Way-like models, we will adopt $R_d = 2.5 \; \textrm{kpc}$ (\citealt{SB09} and references therein) and $l_d = 0.6 \; l_1$ (\citealt{RF12}), implying $R_1 = 8.3 \; \textrm{kpc}$, coincidentally very close to the solar radius. Note that we have chosen values of $R_d$ and $l_d$ that are appropriate to the stellar component alone; this is reasonable for the Milky Way, in which the contribution of the cold gas to the total mass and angular momentum of the disc is negligible. We will also assume that the corona has a size comparable to the virial radius $R_\textrm{vir} \sim 200 \; \textrm{kpc}$ (\citealt{Dehnen+06}), corresponding, in dimensionless units, to $s_\textrm{max} = 24$. We emphasize however that, thanks to our dimensionless formulation, our models can be straightforwardly re-scaled to other choices of the fiducial values, either for the Milky Way or for external galaxies.

A list of the most important dimensionless variables used in this paper is given in Table \ref{table::variables}, for reference. The last three entries will be defined in following sections.

\begin{table}
\centering
\caption{Main dimensionless variables used in this work.}\label{table::variables}
\begin{tabular}{ccc}
\hline
Symbol & Meaning & Defining \\
&& equation \\
\hline
$a$ & Inverse temperature & \eqref{adef} \\
$s$ & Cylindrical radius & \eqref{dimensionlessvariables} \\
$w$ & Rotation velocity & \eqref{wdef} \\
$\lambda$ & Specific angular momentum & \eqref{dimensionlessvariables} \\
$y$ & Equatorial density & \eqref{dimensionlessvariables} \\
$g$ & Inverse AMD & \eqref{gdef} \\
$h$ & Equatorial enthalpy & \eqref{hdef} \\
$\tilde{k}$ & Politrope normalization & \eqref{ktildedef} \\
\hline
\end{tabular}
\end{table}

\subsection{A simple example: coronae with a constant rotation velocity}\label{sec::SimpleTruncatedModels}
To gain some insight, we briefly consider here the simplest possible situation, in which the corona has a constant rotation velocity $V$. This is also useful to give a more quantitative description of some of our qualitative arguments of Section \ref{subsec::preliminary}. From equations \eqref{defpsi}, \eqref{Euler} and \eqref{masseq} and from the requirement that the total mass $M_1$ be finite, a constant rotation velocity also implies a power-law equatorial density distribution $\rho_0$, truncated at a maximum radius $R_\textrm{max}$, and a power law AMD $\psi$, truncated at a maximum specific angular momentum $l_\textrm{max}$. In formulae:
\begin{equation}\label{TruncatedModelDef}
\left.\begin{array}{lll}
V(R) & = & w V_\textrm{d} \quad \forall R\\ [1.em]
\rho_0(R) & = &
\left\{\begin{array}{lll}
\rho_{0, 1} \left( \frac{R}{R_1} \right)^{-q} & \quad & R < R_\textrm{max} \\
0 & \quad & R > R_\textrm{max}
\end{array}\right. \\ [1.em]
\psi(l) & = &
\left\{\begin{array}{lll}
\psi_1 \left( \frac{l}{l_1} \right)^p & \quad & l < l_\textrm{max} \\
0 & \quad & l > l_\textrm{max} \; .
\end{array}\right.
\end{array}\right.
\end{equation}
The shape of the AMD (parametrized by $p$), together with the temperature parameter $a$, completely determines the solution, by defining the value of all the other involved quantities:
\begin{equation}\label{TruncatedModelParameters}
\begin{split}
q & =  2 - p   \\
w  &=  \frac{V}{V_\textrm{d}}  =\sqrt{1 + \frac{p-2}{a}} \\
s_\textrm{max}  &=  \frac{R_\textrm{max}}{R_1}  =  \frac{2+p}{1+p}\frac{1}{w} \\
\lambda_\textrm{max}  &= \frac{l_\textrm{max}}{l_1}  =  \frac{2+p}{1+p} \\
\frac{\rho_{0,1}}{\rho_1} & =  \frac{(1+p)^{2+p}}{(2+p)^{1+p}} w^{1+p} \\
\frac{\psi_1 l_1}{M_1}  & =  \frac{(1+p)^{2+p}}{(2+p)^{1+p}} \; ,
\end{split}
\end{equation}
where we have also made use of the relations \eqref{simplechoice}. Note that $p > -1$ is required by mass convergence at small radii and specific angular momenta. Models with $-1 < p < 2$ have a radially declining equatorial density and sub-centrifugal rotation ($w < 1$). By contrast, models with $p > 2$ have a vanishing central density and a positive radial pressure gradient, leading to vortex-like solutions with super-centrifugal rotation ($w > 1$). The dividing case $p = 2$ corresponds to a corona with uniform equatorial density and in perfect corotation with the disc everywhere. 

From the third equation of \eqref{TruncatedModelParameters} we immediately see that, in this model, a corona with a sensible size ($s_\textrm{max} \gg 1$, see Section \ref{sec::dimensionless}) can only be achieved if its rotation velocity is negligible ($w \ll 1$). However, as we have summarized in Section \ref{subsec::preliminary}, this is in tension both with direct observations and with indirect inference from chemical evolution. Furthermore, $w \ll 1$ would imply $R_\textrm{cross} \gg R_1$ (see equation \ref{Rcrossdef}), which is inconsistent with the requirement that the corona be a source of angular momentum for the disc: if accreting mass from a corona according to equations \eqref{TruncatedModelParameters}, the disc would \emph{decrease} its average specific angular momentum and therefore shrink with time, rather than growing inside-out.

Remarkably, besides any consideration about galaxy evolution, there is an independent reason to discard a model with constant $w$: according to \eqref{TruncatedModelDef}, a constant rotation velocity implies a power-law AMD, which, as we will discuss in more detail in Sections \ref{sec::CosmoAMD} and \ref{sec::WindModels}, is substantially dissimilar to those predicted by cosmological models. We therefore seek a method to build other solutions, appropriate to more general AMDs.

\section{The inversion problem: structure and kinematics from a given angular momentum distribution}\label{sec::inversion}
In order to consider more general situations, we address here the \emph{inversion problem} of reconstructing the structure and the kinematics of the corona from knowledge of its AMD $\psi$. This is a technical section. The reader mostly interested in the applications may wish to skip it and go directly to Section \ref{sec::CosmoAMD}.

Taking the definitions \eqref{defpsi} and \eqref{wdef} into account, equations \eqref{masseq} and \eqref{Euler} can be rewritten as
\begin{equation}\label{newsystem}
\begin{split}
\frac{dl}{dR}(R) & = 4 \pi \chi R^2 \frac{\rho_0(R)}{\psi(l(R))} \\
\frac{d\rho_0}{dR}(R) & = a \frac{\rho_0(R)}{R} \left( \left( \frac{l(R)}{R V_d} \right)^2 - 1 \right) \; .
\end{split}
\end{equation}
The definitions \eqref{dimensionlessvariables} and the dimensionless variable
\begin{equation}\label{gdef}
g(\lambda) \vcentcolon = \frac{M_1}{l_1 \psi(l_1 \lambda)}
\end{equation}
allow the system \eqref{newsystem} to be put in dimensionless form:
\begin{equation}\label{dimensionless}
\begin{split}
\frac{d \lambda}{ds} (s) & = s^2 y(s) g(\lambda(s)) \\
\frac{dy}{ds} (s) & = a \frac{y(s)}{s} \left( \left( \frac{\lambda(s)}{s} \right)^2 - 1 \right) \; .
\end{split}
\end{equation}
 
At least the second of equations \eqref{dimensionless} is singular at the origin, which requires an asymptotic analysis for small values of $s$. Note that part of the singularity is due to the centrifugal barrier $l^2/R^3$, implying that the issue could not be avoided with a different choice of the potential.

We start by isolating the behaviour of the function $g$ for small values of $\lambda$:
\begin{equation}\label{gexpansion}
g(\lambda) = g_1 \lambda^{-p} G(\lambda) \; ,
\end{equation}
with $g_1 > 0$ and $G(0) = 1$. The parameter $p$ is chosen to be consistent with the particular case of Section\ \ref{sec::SimpleTruncatedModels}. Note that the condition $p > -1$ for the convergence of mass holds in the general case, since it only depends on the behaviour of $\psi$ for small values of $l$. 

We then look for a solution of \eqref{dimensionless} in the form
\begin{equation}\label{expansion}
\begin{split}
\lambda(s) & = \lambda_1 s^\alpha A(s) \\
y(s) & = y_1 s^\beta B(s) \; ,
\end{split}
\end{equation}
with $\lambda_1 > 0$, $y_1 > 0$ and $A(0) = B(0) = 1$. Substituting in the first equation of \eqref{dimensionless} and taking the limit for $s \to 0$ one gets
\begin{equation}\label{limitcondition0}
\begin{split}
\beta & = (1 + p)\alpha - 3 \\
y_1 & = \frac{\alpha \lambda_1^{1+p}}{g_1} \; ,
\end{split}
\end{equation}
while from the second equation we have
\begin{equation}\label{limitcondition}
\lambda_1^2 \lim_{s \to 0}s^{2(\alpha-1)} = 1 + \frac{(1+p)\alpha - 3}{a} \; .
\end{equation}

Equation \eqref{limitcondition} can be satisfied for $\alpha \geq 1$. Two branches of solutions arise depending on whether $\alpha = 1$ or $\alpha > 1$.

\subsection{Cold branch}\label{sec::ColdBranch}
If $\alpha = 1$, then from \eqref{limitcondition0} and \eqref{limitcondition}
\begin{equation}
\beta = p-2
\end{equation}
and
\begin{equation}\label{lambda1val}
\lambda_1 = \sqrt{1 + \frac{p-2}{a}}\; ,
\end{equation}
which is well defined for sufficiently large $a$:
\begin{equation}\label{coldbranchainequality}
a > 2 - p \; ,
\end{equation}
i.e., by equation \eqref{adef}, for sufficiently small temperatures, which justifies the name \emph{cold branch} for this class of solutions. 

The expansion \eqref{expansion} then reduces to
\begin{equation}\label{coldbranch}
\begin{split}
\lambda(s) & =  \lambda_1 s A(s) \\
y(s) & = \frac{\lambda_1^{1+p}}{g_1} s^{p-2} B(s) \; ,
\end{split}
\end{equation}
with $A(0) = B(0) = 1$ and $\lambda_1$ given by equation \eqref{lambda1val}. Note that, for small values of $s$, the rotation velocity and the density depend on $p$ as in the toy model of Section \ref{sec::SimpleTruncatedModels} \footnote{From equations \eqref{defw2} and \eqref{coldbranch}, in the cold branch $w(0) = \lambda_1$. \label{foot::lambda1}}. To compute the full structure of the corona we need equations for the functions $A$ and $B$. These are found by inserting equations \eqref{gexpansion} and \eqref{coldbranch} into equations \eqref{dimensionless}:
\begin{equation}\label{ABsystem}
\begin{split}
\frac{dA}{ds}(s) & = \frac{1}{s}\left( B(s)A^{-p}(s)G(\lambda_1sA(s)) - A(s) \right) \\
\frac{dB}{ds}(s) & = (a + p - 2) \frac{B(s)}{s} (A^2(s) -1) \; .
\end{split}
\end{equation}
These equations are also singular at $s = 0$. A solution valid very near to the origin can be found considering the perturbations fields $\delta A$ and $\delta B$ defined by
\begin{equation}\label{perturbationdef} \begin{split}
A \vcentcolon & = 1 + \delta A \\
B \vcentcolon & = 1 + \delta B \; .
\end{split} \end{equation}
To the linear order and for small values of $s$, $\delta A$ and $\delta B$ are governed by the equations
\begin{equation}\label{linearsys}
\begin{split}
\frac{d(\delta A)}{ds} (s) & = G_0^\prime \lambda_1 + \frac{\delta B(s) - (1+p) \delta A(s)}{s} \\
\frac{d(\delta B)}{ds} (s) & = 2 (a+p-2) \frac{\delta A(s)}{s}  \; ,
\end{split} \end{equation}
which, requiring $\delta A (0) = \delta B (0) = 0$, admit the general solution
\begin{equation}\label{generallinearizedsolutioncoldbranch} \begin{split}
\delta A (s) & = A_1 s + A_2 s^\epsilon \\
\delta B (s) & = B_1 s  + B_2 s^\epsilon\; ,
\end{split} \end{equation}
where
\begin{equation} \begin{split}
\epsilon & = \frac{1}{2} \left (\sqrt{(1+p)^2 + 8(a+p-2)} - (1+p) \right) \\
A_1 & = \frac{G^\prime_0 \lambda_1}{6-p-2a} \\
B_1 & = 2(a+p-2) A_1 \\
B_2 & = \frac{2(a+p-2)A_2}{\epsilon} \; 
\end{split} \end{equation}
and $A_2$ is an integration constant. Note that $\epsilon$ is always well defined and positive when the condition \eqref{coldbranchainequality} is met. The solution \eqref{generallinearizedsolutioncoldbranch}, computed at some very small but positive $s \ll 1$, can be used as the initial condition for an ordinary numerical integration of the system \eqref{ABsystem}. The constant $A_2$ can then be fixed \emph{a posteriori} by requiring a boundary condition to be met at large radii.

\subsection{Hot branch}\label{sec::HotBranch}
For $\alpha > 1$, equations \eqref{limitcondition0} and \eqref{limitcondition} give
\begin{equation}\label{tuozio}
a = 3-(1+p)\alpha \;
\end{equation}
and
\begin{equation}
\beta = -a \;.
\end{equation}
Note that inserting $\alpha> 1$ into equation \eqref{tuozio} gives
\begin{equation}\label{hotbrancha}
a  < 2 - p \; ,
\end{equation}
which is complementary to the condition \eqref{coldbranchainequality} and therefore justifies the name \emph{hot branch} for this second class of solutions. The expansion \eqref{expansion} reads
\begin{equation}\label{generalcoldbranch}
\begin{split}
\lambda(s) & =  \lambda_1 s^\alpha A(s) \\
y(s) & = \alpha \frac{\lambda_1^{1+p}}{g_1} s^{-a} B(s) \; ,
\end{split}
\end{equation}
with $A(0) = B(0) = 1$ and \emph{arbitrary} $\lambda_1 > 0$. Note that, in contrast with the cold branch case, the rotation velocity always vanishes in the origin and the inner slope of the density profile is a function of the temperature alone (in particular, it is independent on $p$). The equations for $A$ and $B$ are, in this case,
\begin{equation}\label{hotbranchABsystem}
\begin{split}
\frac{dA}{ds}(s) & = \frac{\alpha}{s}\left( B(s)A^{-p}(s)G(\lambda_1s^{\alpha}A(s)) - A(s) \right) \\
\frac{dB}{ds}(s) & = a \lambda_1^2 s^{2\alpha-3} A^2(s) B(s)\; ,
\end{split}
\end{equation}
with $a$ and $\alpha$ related by equation \eqref{tuozio}. The system \eqref{hotbranchABsystem} is, again, singular at $s  = 0$. Near the origin, the perturbation fields $\delta A$ and $\delta B$, defined as in equation \eqref{perturbationdef}, obey the system
\begin{equation}
\begin{split}
\frac{d(\delta A)}{ds} (s) & = \alpha G_0^\prime \lambda_1 s^{\alpha-1} + \alpha \frac{\delta B(s) - (1+p) \delta A(s)}{s} \\
\frac{d(\delta B)}{ds} (s) & = a \lambda_1^2 s^{2 \alpha -3}  \; .
\end{split}
\end{equation}
The field $\delta B$ is trivially given by
\begin{equation}
\delta B (s) = B_1 s^{2(\alpha-1)} \;
\end{equation}
with
\begin{equation}
B_1 = \frac{a \lambda_1^2}{2(\alpha-1)} \; ,
\end{equation}
while three sub-cases must be considered for $\delta A$:
\begin{equation}
\delta A (s) = \frac{\alpha}{(3+p)\alpha - 2} B_1 s^{2(\alpha -1)}
\end{equation}
if $G_0^\prime = 0$ \emph{or} if $\alpha < 2$ ($a > 1-2p$),
\begin{equation}
\delta A (s) = \frac{G_0^\prime \lambda_1}{2+p} s^\alpha
\end{equation}
if $G_0^\prime \neq 0$, \emph{and} $\alpha > 2$ ($a < 1-2p$) and finally
\begin{equation}
\delta A (s) = \frac{G_0^\prime \lambda_1 + B_1}{2+p} s^2
\end{equation}
if $\alpha = 2$ ($a = 1-2p$).

The full solution is computed with a similar strategy to the cold branch case, with the difference that now $\lambda_1$ (rather than $A_2$) is the integration constant, to be fixed by an outer boundary condition.

\subsection{Remark}\label{sec::Remark}
By equations \eqref{coldbranchainequality} and \eqref{hotbrancha}, the critical value of $a$ (or the critical temperature) marking the transition between the cold and the hot branch is a function of $p$ and can therefore vary, even for a fixed potential, if different AMDs $\psi$ are considered. However, since the integrability of $\psi$ at small $l$ requires $p > -1$, the hot-branch solutions (equation \ref{hotbrancha}) are always super-virial ($a < 3$) and therefore cannot stand in isothermal equilibrium without confinement by an external medium (see Section \ref{sec::starteq}). The cold branch, which contains all the sub-virial and some of the super-virial solutions, is therefore sufficiently general, as long as one is interested in unconfined isothermal solutions with finite mass.

\section{A minimal cosmologically motivated model}\label{sec::CosmoAMD}
Here we apply our reconstruction method to a cosmologically motivated AMD. We consider the class of exponentially truncated power laws, which has been  found to provide a good description for both dark matter and hot gas in cosmological simulations (\citealt{SharmaSteinmetz2005}):
\begin{equation}\label{SS05}
\psi(l) = \psi_{0, p} \left( \frac{l}{l_1} \right)^p e^{- \omega_p l/l_1} \; ,
\end{equation}
where
\begin{equation}\label{SS05param}\begin{split}
\psi_{0, p} & = \frac{(1+p)^{1+p}}{\Gamma(1+p)} \frac{M_1}{l_1} \\
\omega_p & = 1 + p
\end{split}\end{equation}
and $\Gamma$ is the complete gamma function. For small values of $l$, the distribution \eqref{SS05} is similar to that of Section \ref{sec::SimpleTruncatedModels}, but at large $l$ it has a gradual rather than a sharp truncation.

It must be pointed out that the simulations of \cite{SharmaSteinmetz2005} were \emph{non-radiative}. While neglecting radiative processes is obviously unreasonable to study galaxy formation, this approach gives the cleanest prediction of the outcome of \emph{gravitational} processes -- in particular, tidal torques -- which are primarily responsible for the shape of the AMD of structures. The distribution \eqref{SS05} should therefore be considered as a first order description for a dark matter halo, as well as for the baryons that have yet to experience significant cooling (essentially, our definition of a cosmological corona). Non-gravitational processes can nonetheless have an impact on the AMD of the hot halo and in particular to increase the average specific angular momentum of the hot gas with respect to its primordial value (e.g.\ \citealt{MallerDekel2002}); for this reason we will also consider, in Section \ref{sec::WindModels}, strong modifications to the `primordial' AMD above.

\cite{SharmaSteinmetz2005} also found the hot gas to have a slightly larger average specific angular momentum $l_1$ with respect to the dark matter, even in the absence of non-gravitational processes. This result is, however, more controversial (e.g. \citealt{vdB+02}; \citealt{Chen+03}) and still lacks a clear physical explanation. In more recent non-radiative simulations, \cite{Sharma+12} found the ratio between the average specific angular momenta of the hot halo and of the dark matter to oscillate with time, in response to merger stages, with a moderate excursion (typically 20 \%) around unity. In the rest of this section, we will assume that $l_1$ is equal to the average specific angular momentum of the dark matter. We recall, however, that, as for all our fiducial values, the effect of other choices can always be drawn out from our models straightforwardly by re-scaling (see Section \ref{sec::dimensionless}).

We now focus on equation \eqref{SS05} and in particular on the elementary case $p = 0$, which is among the most common values found in the simulations by \cite{SharmaSteinmetz2005} and \cite{Sharma+12} and is also in agreement with the asymptotic behaviour of $\psi$, for small $l$, found in the dark-matter-only simulations of \cite{Bullock+01}. For $p = 0$, equations \eqref{SS05}-\eqref{SS05param} are simply reduced to the exponential AMD:
\begin{equation}\label{ExpModel}
\psi(l) = \frac{M_1}{l_1} e^{-l/l_1} \; .
\end{equation}

In principle, the method described in Section \ref{sec::inversion} can be applied to the distribution \eqref{ExpModel}, fixing the integration constant to meet the boundary condition $l (R) \to + \infty$ for $R \to \infty$. This would be, however, complicated in practice (for our method requires a necessarily finite numerical integration) and not really physically meaningful (because real structures will always have a maximum specific angular momentum). For our practical application, we therefore consider a modified version of \eqref{ExpModel}, to account for a maximum value of $l$.

The general truncated exponential AMD reads
\begin{equation}\label{truncatedexp}
\left.\begin{array}{lll}
\psi(l) & = &
\left\{\begin{array}{lll}
\psi_0 e^{- \omega \frac{l}{l_1}} & \quad & l < l_\textrm{max} \\ [1.em]
0 & \quad & l > l_\textrm{max} \; ,
\end{array}\right.
\end{array}\right.
\end{equation}
where the three parameters $\psi_0$, $\omega$ and $l_\textrm{max}$ are subject to the two constraints that the total mass is $M_1$ and the average specific angular momentum is $l_1$ (equation \ref{simplechoice}), leaving just one degree of freedom. The one-dimensional family of consistent models is conveniently parametrized by the quantity $\xi \vcentcolon = \omega l_\textrm{max}/l_1$, in terms of which, imposing the conditions \eqref{simplechoice}, the three parameters can be written as
\begin{equation}\label{csiparam}
\begin{split}
\omega & = 1 - \frac{\xi e^{-\xi}}{1 - e^{-\xi}} \\
\lambda_\textrm{max} \vcentcolon = \frac{l_\textrm{max}}{l_1} & =  \frac{\xi(1  -e^{-\xi})}{1-(1+\xi)e^{-\xi}}\\
P \vcentcolon= \frac{\psi_0 l_1}{M_1} & = \frac{1-(1+\xi)e^{-\xi}}{(1-e^{-\xi})^2} \; .
\end{split}
\end{equation}
The parameter $\xi$ can assume all non-negative real values. When $\xi$ varies from 0 to $+\infty$, $\omega$ increases from 0 to 1, $\lambda_\textrm{max}$ from  2 to $+ \infty$, and $P$ from $1/2$ to $1$. Note that the limits $\xi \to 0$ and $\xi \to +\infty$ correspond, respectively, to the uniform distribution (the case $p = 0$ in Section \ref{sec::SimpleTruncatedModels}) and to the non-truncated exponential \eqref{ExpModel}, as expected.

\begin{figure*}
\centering
\includegraphics[width=18cm]{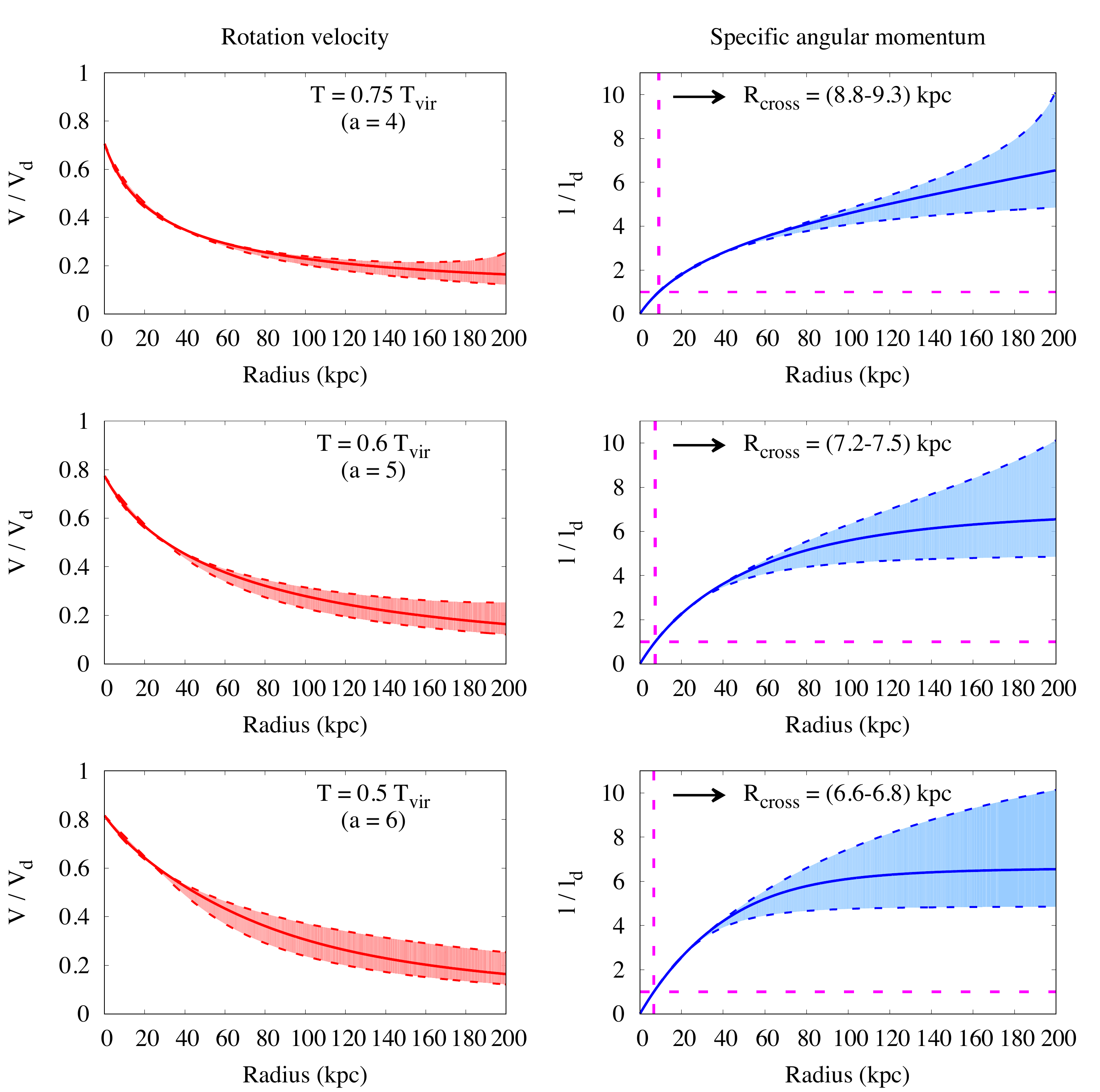}
\caption{Kinematic properties of a sub-virial isothermal corona with an exponential AMD. Radial profiles are given for the rotation velocity (left), in units of the velocity of the disc $V_d$, and for the specific angular momentum (right), in units of the average specific angular momentum of the disc $l_d$, for three sub-virial values of the temperature parameter $a = 4$ (top), $a = 5$ (middle) and $a = 6$ (bottom). Shaded areas enclose the dependence on boundary conditions, adopting the average (solid line) and $1 \sigma$ scatter in maximum specific angular momentum from \citet{Bullock+01}. In the right-hand panels, the horizontal and vertical dashed lines and the labels highlight the radius $R_\textrm{cross}$ out of which coronal gas with $l > l_d$ is stored. In these models, the rotation velocity declines with radius, from moderately sub-centrifugal ($V \lesssim V_d$) on galactic scales to dynamically negligible ($V \ll V_d$) close to the virial radius. At low $R$, the specific angular momentum rises steeply with radius, so that the coronal gas with $l > l_{d}$, which is needed to sustain inside-out growth, is in direct contact with the disc and available for accretion. \label{fig::Fig1}}
\end{figure*}

We consider here three values $\xi = 2.0$, $\xi = 3.5$ and $\xi = 6.0$, corresponding, respectively, to $\lambda_\textrm{max} = 2.9$, $\lambda_\textrm{max} = 3.9$ and $\lambda_\textrm{max} = 5.9$, which are the average and $1 \sigma$ interval found by \cite{Bullock+01}. Following the fiducial scaling of Section \ref{sec::dimensionless}, we choose $s_\textrm{max} = 24$, corresponding to $R_\textrm{max} = 200 \; \textrm{kpc}$. Note that, since the specific angular momentum is always an increasing function of radius, the effect of changing $s_\textrm{max}$ at fixed $\lambda_\textrm{max}$ is qualitatively similar to that (explicitly explored here) of changing $\lambda_\textrm{max}$ at fixed $s_\textrm{max}$. For the average angular momentum of the disc, we adopt the fiducial scaling $l_d = 0.6 \; l_1$ inferred by \cite{RF12}, as also described in Section \ref{sec::dimensionless}.

\subsection{Kinematics}
In Figure \ref{fig::Fig1}, the rotation velocity and specific angular momentum profiles are shown for three sub-virial values of the temperature parameter $a > 3$. 

With a small dependence on the temperature and on $\lambda_\textrm{max}$, the kinematic properties of these models are similar. In the inner regions, the rotation velocity is lower than, but comparable to, the velocity of the disc and then it gently declines to much lower values at large radii. This is in agreement with the expected behaviour (see Section \ref{subsec::preliminary}). The inner rotation is between 70 and 80 \% of the rotation velocity of the disc, as suggested by chemical evolution models of the Milky Way (\citealt{BS12}; \citealt{PF16}). Assuming $V_d = 240 \; \textrm{km} \; \textrm{s}^{-1}$ (\citealt{Schoenrich2012}), this corresponds to $V = 168-192 \; \textrm{km}\;\textrm{s}^{-1}$, in very good agreement with the observational estimate of \cite{HB16}. The specific angular momentum increases steeply in the inner regions and more gently in the outskirts. The steep inner rise allows $l$ to become larger than the disc average $l_d$ already at quite small radii $s \sim 1$, or, in our Milky Way scaling, at about the solar radius (see Section \ref{sec::dimensionless}). This is quantified by $R_\textrm{cross}$, as defined in equation \eqref{Rcrossdef} and shown in Figure \ref{fig::Fig1}. As explained in Section \ref{subsec::preliminary}, the important consequence of the described behaviour is that, in these models, the disc is directly in contact with coronal material that is rich enough in angular momentum to be a source for the inside-out growth of the disc. Note that the inner parts of the angular momentum profiles, on which the conclusion above is based, are fairly insensitive to the choice of the boundary condition, as shown in Figure \ref{fig::Fig1} by the shaded regions and by the narrow range of $R_\textrm{cross}$ at fixed temperature. We also point out that the decrease of $R_\textrm{cross}$ with decreasing temperature is related to a corresponding increase of the central rotation velocity, which was expected already on the basis of the simple estimate in equation \eqref{V0_LowerLimit}. We emphasize that the match between model behaviour and expectations, obtained here under minimal assumptions, is not obvious: we will see, in the next Sections, that the same agreement is not easily recovered -- and sometimes completely lost -- when more complicated situations are considered.

\subsection{Density distribution}\label{sec::density}

\begin{figure*}
\centering
\includegraphics[width=8.5cm]{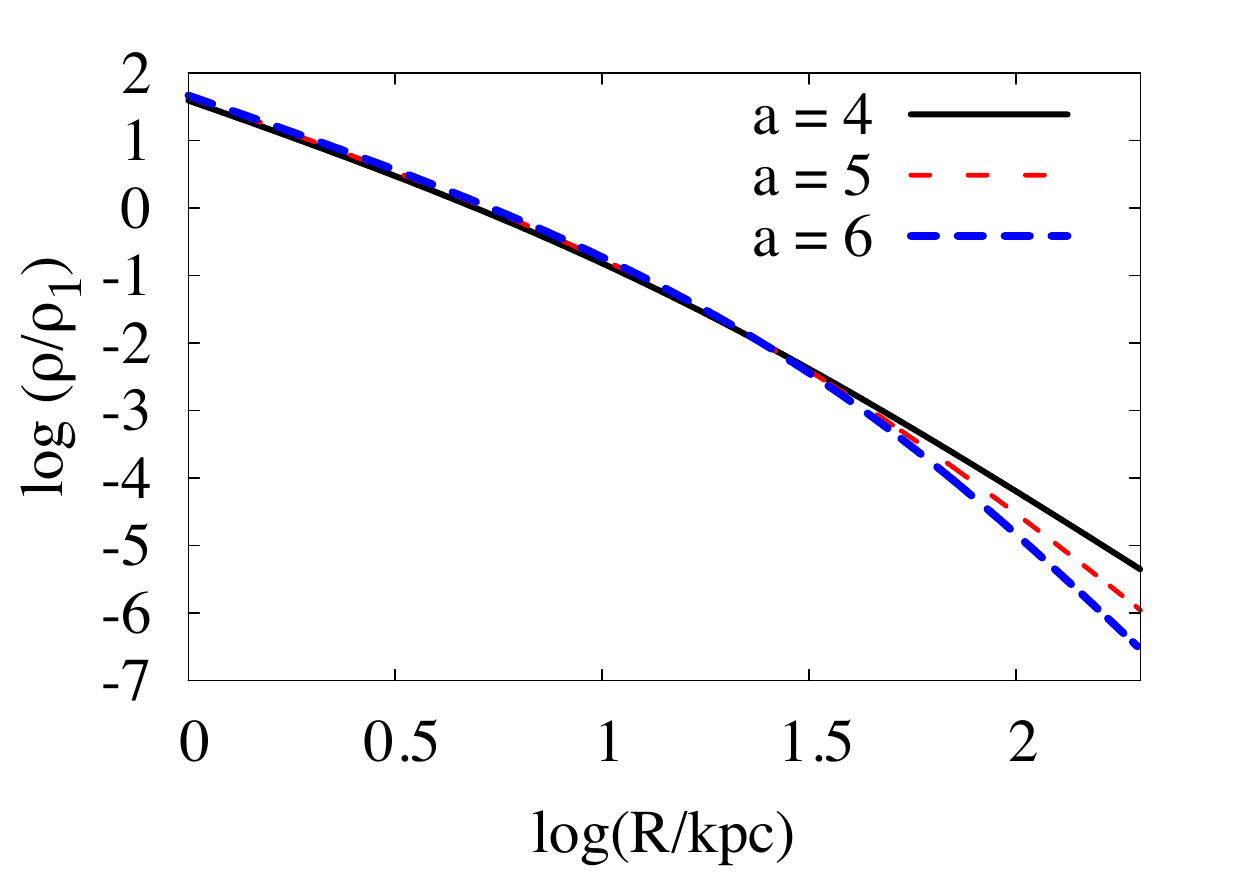}
\caption{Equatorial density profile for different values of the temperature parameter $a$. The change of slope at large radii ensures a finite total mass, which is a property of sub-virial models. Note that the scaling density $\rho_1$ is lower for models with higher temperature (see text).}\label{fig::EqDensityProfile}
\end{figure*}

\begin{figure*}
\centering
\includegraphics[width=8.5cm]{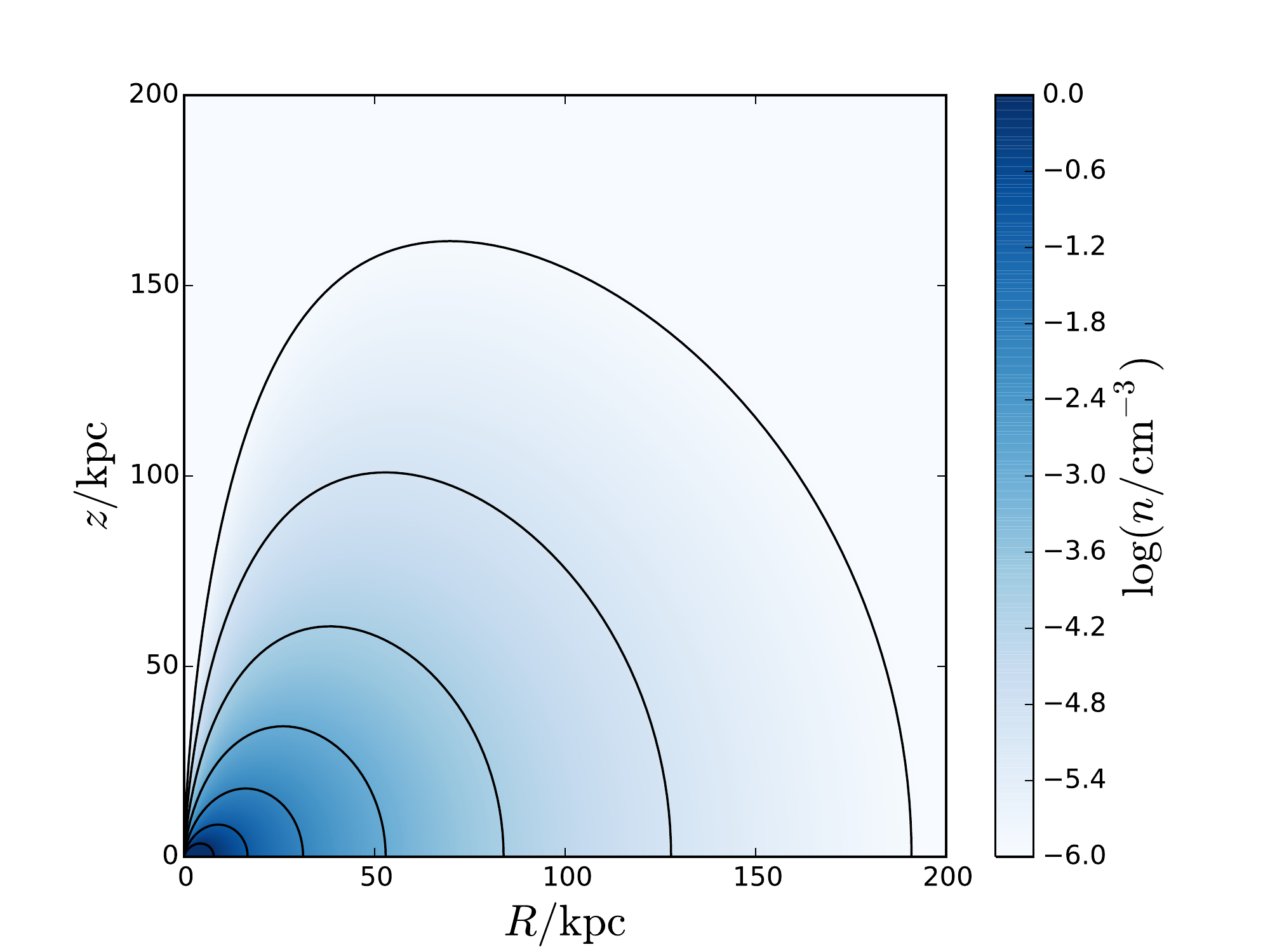}
\caption{Two-dimensional map of the density in the meridional plane for the model with $a = 6$ and $\lambda_\textrm{max} = 3.9$. The density normalization is chosen so that the total mass equals the maximum theoretical value (see text). Contours are shown in steps of 1 dex, the outermost being for a particle number density $n = 10^{-6} \; \textrm{cm}^{-3}$. The prominent depression near the rotation axis is related to the assumption of barotropic equilibrium (see Section \ref{sec::DiscussionBarotropic}).}\label{fig::IsoModelContours}
\end{figure*}

Figure \ref{fig::EqDensityProfile} shows the radial profile of the dimensionless equatorial density $y = \rho_0/\rho_1$ for different values of the temperature parameter $a$ (the dependence on $\lambda_\textrm{max}$ is negligible). All the models follow, in the inner parts, the same power-law $\rho_0 \; \propto \; R^{-2}$, as implied by equation \eqref{coldbranch} with $p = 0$. Note that, if extrapolated indefinitely, this slope would correspond to an infinite total mass (equation \ref{masseq}). However, the profile naturally steepens at large radii in an integrable way, with a slope that is a function of the temperature. This is readily understood from equation \eqref{Euler}: considering the decay of $w$ with radius (Figure \ref{fig::Fig1}), the asymptotic behaviour of the equatorial density for large radii is $\rho_0 \; \propto \; R^{-a}$. Note that the common normalization of the profiles in Figure \ref{fig::EqDensityProfile} is only apparent, since the density scaling factor $\rho_1$ is a decreasing function of the 
temperature (cfr.\ equations \ref{scalings} and \ref{defk}). The (trivial) consequence is that higher temperature coronae are more diffuse.

\begin{figure*}
\centering
\includegraphics[width=18cm]{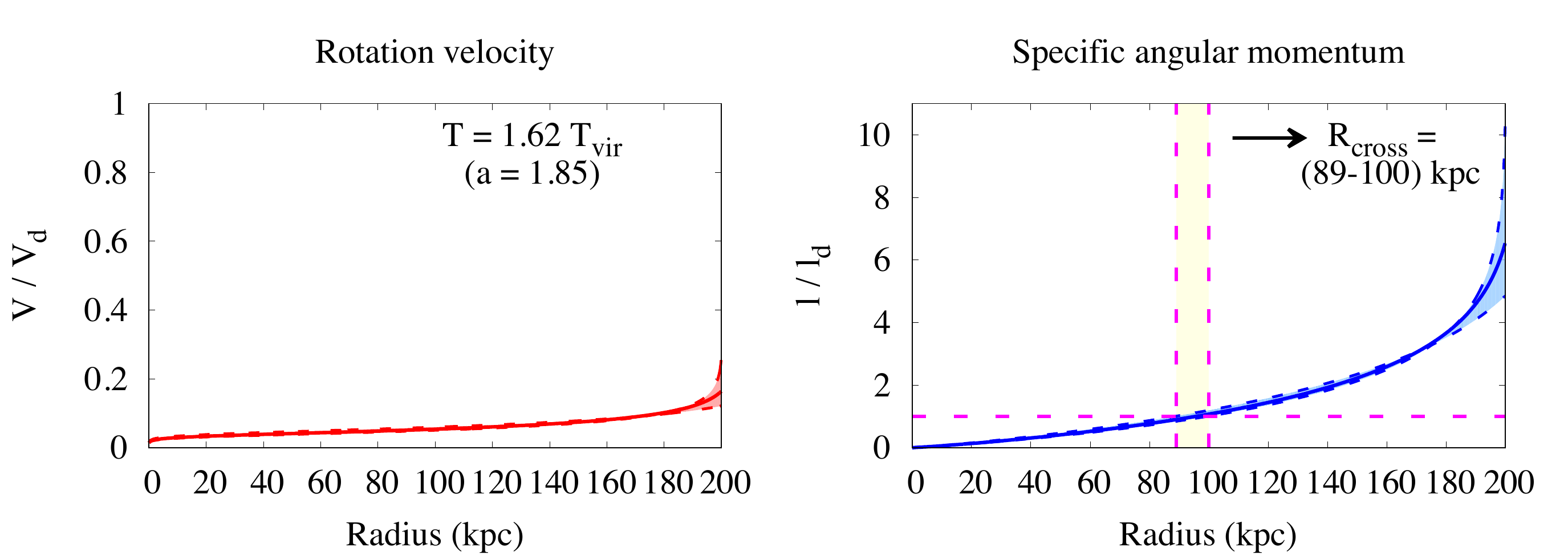}
\caption{Similar to Figure \ref{fig::Fig1}, but for a super-virial value of the temperature parameter $a = 1.85$, suggested for the corona of the Milky Way by X-ray observations. In this model, the rotation velocity is dynamically negligible everywhere. The specific angular momentum rises very slowly with radius, so that the material with $l > l_\textrm{d}$, which would be needed to sustain inside-out growth, is stored far away from the disc and is not available for accretion. The vertical shaded region in the right-hand panel highlights a larger dependence of $R_\textrm{cross}$ on boundary conditions, compared to the sub-virial case.\label{fig::SuperVirial}}
\end{figure*}

In Figure \ref{fig::IsoModelContours}, the two-dimensional map of the density is shown for the case $a = 6$ ($T = 0.5 \; T_\textrm{vir}$) and $\lambda_\textrm{max} = 3.9$, the vertical density decline being computed by means of equation \eqref{verticaleq}. Note that this model has significant deviations from spherical symmetry, although, as we will discuss in Section \ref{sec::DiscussionBarotropic}, the most prominent feature (the very low density funnel near the rotation axis) may disappear if less restrictive assumptions on the thermodynamic state of the gas are made. The predictions close to the equator are also likely inaccurate, since at small heights $|z|$ the cosmological corona will be overwhelmed by material belonging to the disc, or ejected by the disc as a consequence of multiple supernova explosions (e.g.\ \citealt{Melioli+09}; \citealt{Marasco+15}). The most informative region of the plane is therefore the one at intermediate heights and cylindrical radii. To allow for a comparison with observations, we use here physical, rather than dimensionless, units for the density. According to equation \eqref{scalings}, this requires an assumption on the total mass of the corona $M_1$. In this plot, we assume that the halo of the Milky Way has a mass $M_\textrm{halo} = 1.5 \times 10^{12} \; \textrm{M}_\Sun$ (\citealt{Piffl+14} and references therein), that the \emph{cosmic baryon fraction} is $\Omega_\textrm{b}/\Omega_\textrm{m}= 0.155$ (\citealt{Planck2014}) and that the corona contains a fraction $f_\textrm{hot} = 0.7$ of the total theoretical baryonic budget: $M_1 = f_\textrm{hot} f_\textrm{bar} M_\textrm{halo}$. This is the maximum possible mass of the corona, if we allow for 30 \% of the baryons in the form of stars and cold gas, as appropriate for Milky-Way-sized galaxies (\citealt{Dutton+10}). Obviously, different choices of $f_\textrm{hot}$ or $M_\textrm{halo}$ would affect the model densities in proportion. With the adopted scaling, the total particle number density, at a galactocentric distance of 70 kpc, is $n \sim 10^{-4} \; \textrm{cm}^{-3}$, in agreement with the estimate of \cite{Gatto+13} based on ram-pressure stripping of satellite galaxies. At the distance and latitude of the Large Magellanic Cloud (LMC) ($50$ kpc, $33^\degree$) the particle density is $n \sim 10^{-3} \; \textrm{cm}^{-3}$, similar to the \emph{line-of-sight average} density inferred from LMC pulsar dispersion measures (\citealt{AndersonBregman2010} and references therein). Since the model density increases approaching the disc, this value is in tension with observations, though a more quantitative assessment of this point would require an appropriate model of the disc-corona interaction at small $|z|$, which is difficult to achieve for the reasons mentioned above.

\section{A super-virial model}\label{sec::SuperVirialModels}

Until now, we have disregarded models with $T \geq T_\textrm{vir}$ because they violate the virial theorem in isolation (see footnote \ref{foot::virial}). However, recent X-ray observations (\citealt{HenleyShelton2013}) indicate that the Milky Way is surrounded by hot gas with a temperature $T = (2.22 \pm 0.63) \times 10^6 \; \textrm{K}$, significantly larger than the virial temperature $T_\textrm{vir} = 1.37 \times 10^6 \; \textrm{K}$, estimated from equation \eqref{Tvirdef} assuming $V_d = 240 \; \textrm{km} \; \textrm{s}^{-1}$ (\citealt{Schoenrich2012}). This suggests that the corona of the Milky Way may be super-virial, at least in the innermost regions, where the density is higher and most of the X-ray emission originates (see also \citealt{FP06}), and justifies considering models with $T \geq T_\textrm{vir}$ ($a \leq 3$). Here we focus on the case $T = 1.62 \; T_\textrm{vir} \; (a = 1.85)$, which corresponds to the median observed temperature of the corona of the Milky Way. \footnote{The reader interested in technical aspects may notice that with $p = 0$, as appropriate to the AMD \eqref{truncatedexp}, models in the range $2 < a < 3$ ($T_\textrm{vir} < T < 1.5 \; T_\textrm{vir}$) fall into the cold branch (Section \ref{sec::ColdBranch}) and have a finite central rotation velocity $V(0)$, though smaller than in the sub-virial case $a > 3$ (see equation \ref{lambda1val} and footnote \ref{foot::lambda1}). The value $a = 1.85$ considered here falls into the hot branch and can be treated as described in Section \ref{sec::HotBranch}.}

The results of the calculation, reported in Figure \ref{fig::SuperVirial}, are very different to the sub-virial models of Figure \ref{fig::Fig1}. The rotation velocity now vanishes in the origin and increases very slowly with radius, remaining smaller than $0.2 \; V_d$ virtually everywhere. This model is not substantially different from a model with a constant and very low rotation velocity, which, as discussed in Sections \ref{subsec::preliminary} and \ref{sec::SimpleTruncatedModels}, is unable to sustain the inside-out growth of a galaxy disc. This is most clearly seen in the right-hand panel: the specific angular momentum increases with radius much more slowly than in the sub-virial case. As a consequence, it becomes larger than the average specific angular momentum of the disc $l_\textrm{d}$ only at a radius of  $90-100 \; \textrm{kpc}$, too far away to be accessible for direct accretion on to the disc.

In this model, the rotation velocity is dynamically negligible everywhere. As a consequence, the density distribution (not shown) is similar to the classical, spherically symmetric, distribution usually assumed by non-rotating models of galactic coronae. According to equation \eqref{Euler}, the density slope is almost constant $d \ln \rho_0 / d \ln R \simeq -a = -1.85$. This value is intermediate between the slope $(-2)$ of our sub-virial models (Section \ref{sec::density}) and the slope $(-1.5)$ of the recent non-rotating model of \cite{MillerBregman2015}.

The failure of this model in storing significant angular momentum in the inner regions can be easily understood. Increasing the temperature has the effect of progressively moving mass towards the outer regions of the corona. For a fixed AMD, the angular momentum is bound to follow the mass, with the result that the bulk of the angular momentum will also be stored far away from the disc and become inaccessible to accretion. In the next two sections, we investigate whether a corona as hot as observed for the Milky Way can be made compatible with being a driver of inside-out growth, by relaxing some of our assumptions, namely a `primordial' AMD (Section \ref{sec::WindModels}) and isothermality (Section \ref{sec::PoliModels}).

\section{The effect of a Galactic wind}\label{sec::WindModels}
\begin{figure*}
\centering
\includegraphics[width=18cm]{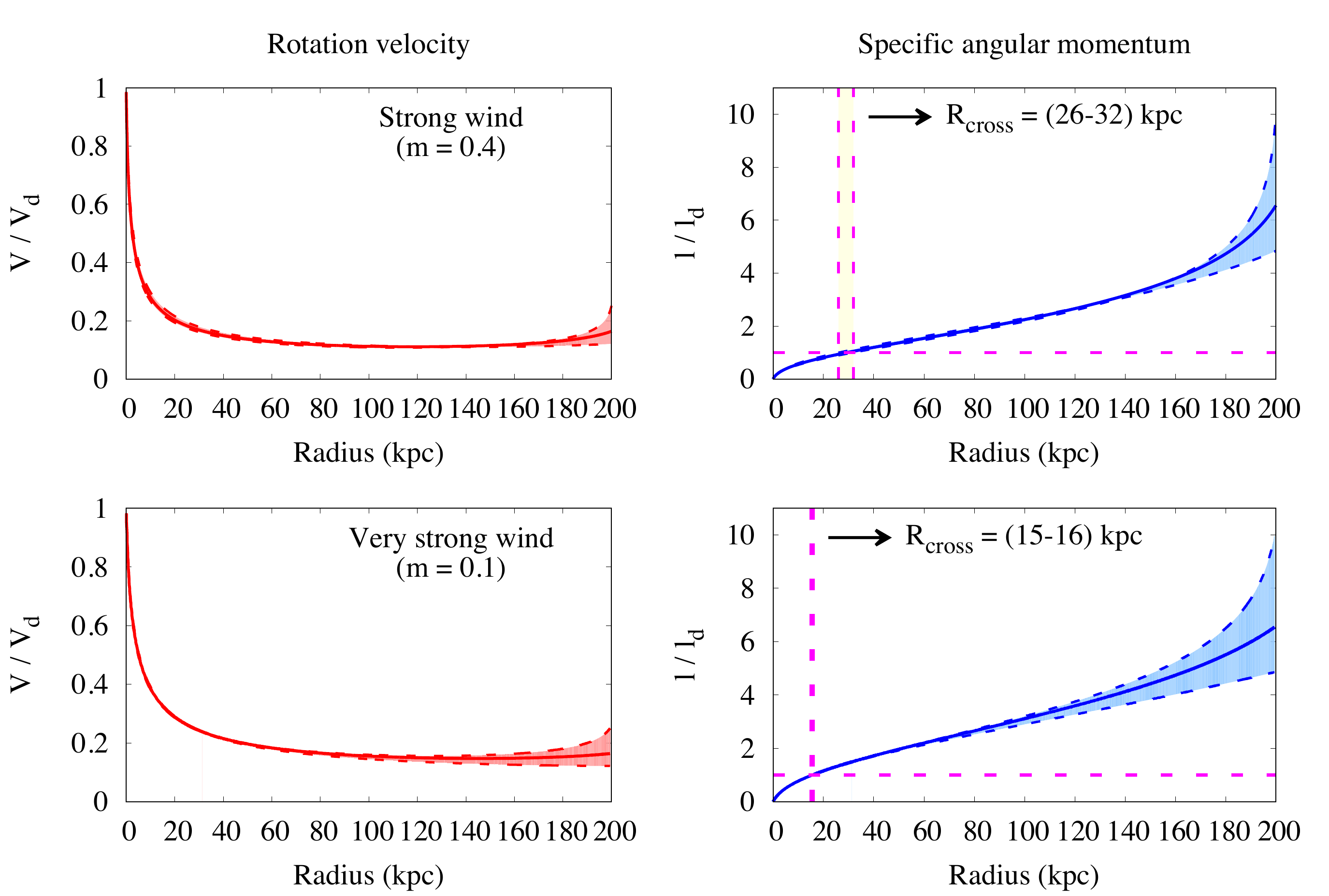}
\caption{Similar to Figure \ref{fig::SuperVirial}, but assuming that a Galactic wind expelled the low angular momentum material from the halo, leaving a \emph{surviving} mass equal to a fraction $m = 0.4$ (upper panels) or $m = 0.1$ (lower panels) of the initial value. Note that in these models the average specific angular momentum of the corona is larger than $l_1$ (see text). The rotation velocity is high in the centre, but declines very steeply with radius. The angular momentum rises relatively slowly and even the model with the most extreme feedback ($m = 0.1$) is only marginally compatible with driving inside-out growth, since $l$ becomes larger than $l_\textrm{d}$ only at the edge of the Galactic disc. \label{fig::Wind}}
\end{figure*}

In Sections \ref{sec::CosmoAMD} and \ref{sec::SuperVirialModels} we have seen that gas accretion from a sub-virial corona with a cosmologically motivated AMD can sustain the inside-out growth of a Milky-Way-like spiral galaxy, while a corona with a super-virial temperature cannot, for it would only be able to provide material with too small angular momentum. It can be argued, however, that inside-out growth of the disc could be achieved also in the super-virial case, if a different AMD, richer in high angular momentum gas, is considered.

The problem of excessive accretion of low angular momentum material is well known from numerical studies of galaxy formation in a cosmological context (e.g.\ \citealt{Navarro+97}). A very popular solution is to invoke strong galactic winds, powered by star formation or a central active galactic nucleus (AGN), to expel large masses of gas from the halo (e.g.\ \citealt{Governato+10}). Crucially, if the material removed from the halo preferentially comes from the innermost parts, where the gas has the lowest angular momentum, the remaining material will have a significantly modified AMD and, in particular, a larger average specific angular momentum (\citealt{MallerDekel2002}; \citealt{Brook+11}). The disappearance of the inner low-angular momentum gas will cause the remaining gas, with large angular momentum, to fall towards the central regions, coming in contact with the disc where it could be accreted to sustain inside-out growth. If angular momentum is conserved in the process, the infalling material would also increase its rotation velocity, producing an inner fast-spinning corona, as suggested by recent observations and required by chemical evolution constraints (see Section \ref{subsec::preliminary}).

To model the preferential expulsion of low angular momentum gas, we multiply our `primordial' AMD \eqref{truncatedexp} by a modulating function $\theta$, which suppresses the low-angular momentum part of the distribution, while leaving (asymptotically) unaltered the high angular momentum part:
\begin{equation}\label{windpsi}\begin{split}
\psi_{\theta} & = \theta \psi \\
\theta (\lambda) & = \frac{\left(\frac{\lambda}{\lambda_0}\right)^p}{1 + \left(\frac{\lambda}{\lambda_0}\right)^p} \; ,
\end{split}\end{equation}
with $p > 0$ and $\lambda_0 > 0$.

To model the effect of feedback in the most favourable conditions for inside-out growth, we consider here extreme values of the parameters. We choose $p = 2$, which forces the central rotation velocity to be as large as the centrifugal speed $w(0) = \lambda_1 = 1$. \footnote{Note that, with $p = 2$, the super-virial value $a = 1.85$ falls in the cold branch, so that equation \eqref{lambda1val} applies (see Section \ref{sec::Remark}).} We then fix $\lambda_0$ by requiring the residual mass (the integral of $\psi_\theta$ in \ref{windpsi}) be some fraction $m$ of the original mass $M_1$ (the integral of $\psi$). We consider two values $m = 0.4$ and $m = 0.1$. If 30 \% of the original baryonic mass is condensed into stars and cold gas (\citealt{Dutton+10}), these values correspond to the disc having ejected from the halo, in the form of winds, an amount of gas equal to, or double, its own mass, respectively. A residual coronal mass equal to 10 \% of the total baryonic budget is also the lowest observational estimate derived by current (non-rotating) models of the corona of the Milky Way (e.g.\ \citealt{MillerBregman2015}) and can therefore be considered as an extreme case. The average specific angular momentum of the surviving part of the corona is $\sim 1.6 \; l_1$ for $m = 0.4$ and $\sim 2 l_1$ for $m = 0.1$, with a negligible dependence on $\lambda_\textrm{max}$. The extreme case, in which the hot gas has twice the specific angular momentum of the dark matter, is in good agreement with recent cosmological simulations including feedback in the form of massive, high (larger than escape) velocity winds (\citealt{Teklu+15}).

The results of this model, with the same value $a = 1.85$ estimated in Section \ref{sec::SuperVirialModels}, are given in Figure \ref{fig::Wind}. The rotation velocity is (by construction) very high in the centre, but falls off very rapidly with radius. As a consequence, the  inner angular momentum profile is steeper than in the super-virial model without winds (Figure \ref{fig::SuperVirial}) but still not as steep as in the minimal sub-virial models (Figure \ref{fig::Fig1}). As expected, the removal of low $l$ material enhances the ability of the corona to provide the disc with fresh high angular momentum gas. However, only in the most extreme model ($m = 0.1$) does the angular momentum rise rapidly enough to become larger than the disc average at reasonably small radii and, even in the extreme case, this happens only at a radius of $\sim 15 \; \textrm{kpc}$, close to the outer edge of the Galactic disc. We conclude that, even when pushing the effect of mechanical feedback to an extreme, an isothermal corona with $T \simeq 2 \; \times 10^6 \; \textrm{K}$ is, at most, marginally consistent with being able to sustain inside-out growth.

We emphasize that the negative conclusion above is specific to models with a large temperature. As we have seen in Section \ref{sec::CosmoAMD}, sub-virial models easily succeed in matching the inside-out growth requirement already with a primordial AMD ($m = 1$, in the present formalism) and any modification of the type \eqref{windpsi} (due to a wind, or simply to the fact that the disc condensed out of relatively low $l$ material) would further enhance the ability of the corona to sustain the inside-out growth of the disc.

\section{The effect of a temperature gradient}\label{sec::PoliModels}
We investigate here the effect of relaxing our assumption that the corona is exactly isothermal. As already noted in Section \ref{sec::SuperVirialModels}, observational determinations of the temperature of the Galactic corona are dominated by the signal coming from the innermost region and do not rule out the existence of a temperature gradient in the radial direction. It is therefore possible that only the inner parts of the corona have been heated to a super-virial temperature, maybe as a consequence of feedback. In this view, a model with a temperature gradient might be considered as an (equally extreme) alternative to the model of Section \ref{sec::WindModels}: rather than being mechanically advected out of the halo in the form of a wind, the energy released by feedback processes may be deposited into the inner parts of the corona in the form of thermal energy and then gradually radiated away, leaving the AMD unaltered. Note also that, in contrast with the isothermal super-virial case, a model with a temperature gradient does not need, in general, to be confined by an external medium.

To model a temperature gradient, we assume that the coronal gas obeys a relation of the form
\begin{equation}\label{polidef}
P = k  \rho^\gamma
\end{equation}
with $k > 0$ and $1 < \gamma \leq 5/3$. The case $\gamma = 5/3$ is for an isentropic configuration (and in this case the constant $k$ is a function of the specific entropy), $\gamma = 1$ (excluded here) would be appropriate to the isothermal case, while $\gamma > 5/3$ would make the models prone to convective instability (e.g.\ \citealt{Tassoul2000}). Note that equation \eqref{polidef} is still a barotropic prescription (i.e.\ with density and pressure stratified on the same surfaces) and therefore implies (as in the isothermal case) that the rotation velocity $V$ is a function of the cylindrical radius $R$ alone. The more general case of baroclinic rotating equilibria (see e.g.\ \citealt{Barnabe+06}) is left for future investigations.

\begin{figure}
\centering
\includegraphics[width=8.5cm]{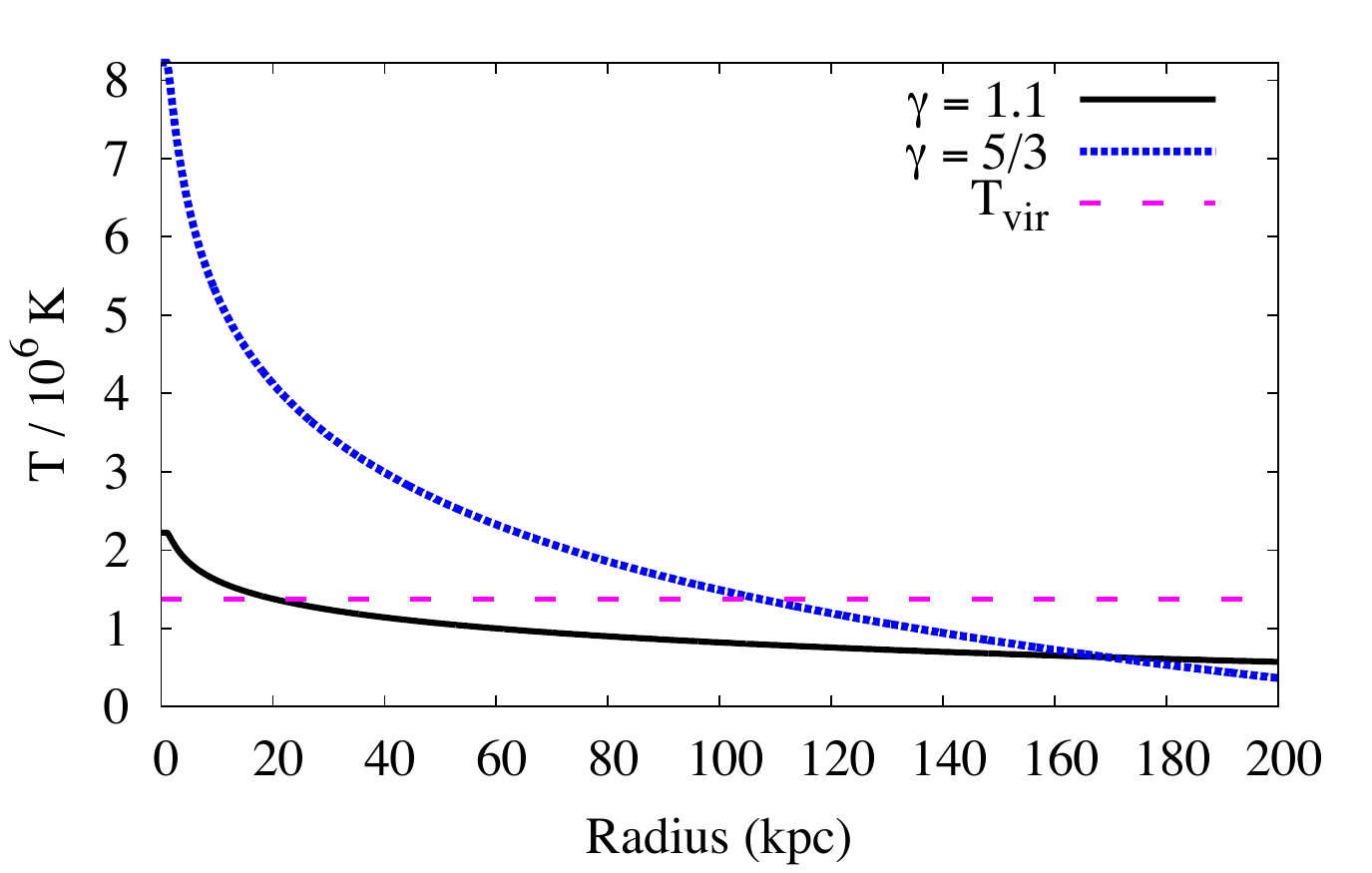}
\caption{Equatorial temperature profile for two very different politropic models: close-to-isothermal ($\gamma = 1.1$, solid black line) and isentropic ($\gamma = 5/3$, dotted blue line). The horizontal dashed line is the virial temperature of the Milky Way, according to the definition \eqref{Tvirdef}.
\label{fig::Tplot}}
\end{figure}

\begin{figure*}
\centering
\includegraphics[width=18cm]{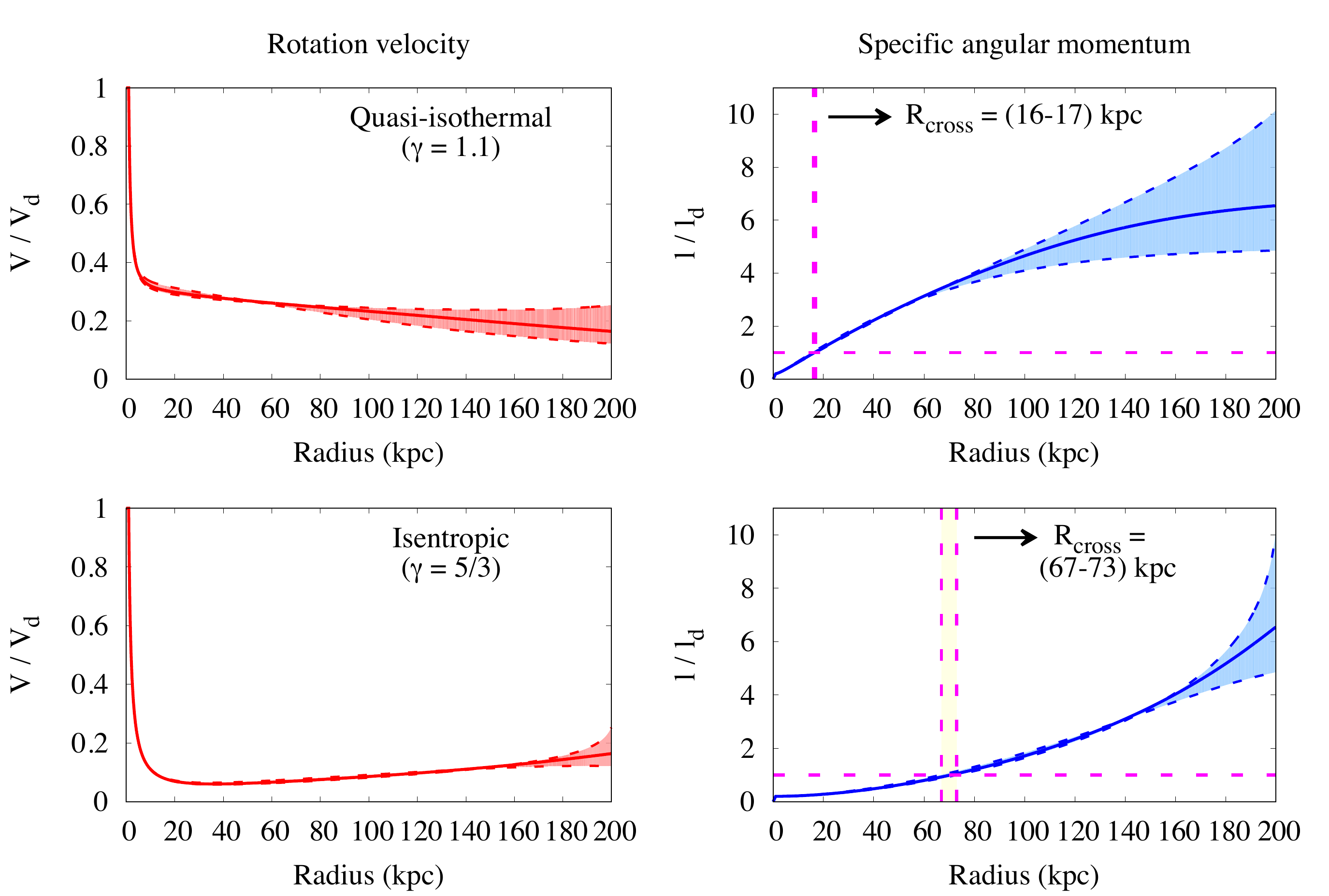}
\caption{Similar to Figure \ref{fig::Fig1}, but assuming a temperature gradient as in Figure \ref{fig::Tplot}. The close-to-isothermal model ($\gamma = 1.1$) is similar to the extreme wind model ($m = 0.1$ in Figure \ref{fig::Wind}) in being marginally consistent with inside-out growth. In the isentropic model ($\gamma = 5/3$), instead, the high-$l$ coronal gas is far away from the disc.\label{fig::Politropic}}
\end{figure*}

Equation \eqref{verticaleq} for the vertical density profile is replaced by
\begin{equation}\label{density2}
\rho (R, z) = \rho_0(R) \left( 1 - \frac{1}{2 h(R)} \ln \left( 1 + \left( \frac{z}{R} \right)^2 \right) \right)^{\frac{1}{\gamma-1}} \; ,
\end{equation}
where
\begin{equation}\label{hdef}
h(R) \vcentcolon = \frac{k \gamma}{\gamma-1}\frac{(\rho_0(R))^{\gamma-1}}{V_d^2}
\end{equation}
is the dimensionless equatorial specific enthalpy (we use this name for simplicity, although it would be strictly appropriate only in the isentropic case). Note that the enthalpy $h$ is proportional to the temperature $T$:
\begin{equation}
h =  \frac{\gamma}{3(\gamma-1)}\frac{T}{T_\textrm{vir}} \; .
\end{equation}
Note also that the density vanishes (together with its first vertical derivative), at a maximum height
\begin{equation}\label{zmax}
z_\textrm{max} (R) = R \sqrt{ e^{2h(R)} - 1 } \; .
\end{equation}
The projected surface density profile is given by
\begin{equation}
\Sigma(R) = 2 \int_0^{z_\textrm{max(R)}} \rho(R,z) dz = 2 R \rho_0(R) F(h(R)) \; ,
\end{equation}
where
\begin{equation}
F(h) \vcentcolon = h \int_0^1 x^\frac{1}{\gamma-1} \frac{e^{2h(1-x)}}{\sqrt{e^{2h(1-x)}-1}} dx \; .
\end{equation}

The radial equation of hydrodynamic equilibrium in the equatorial plane is easily written in terms of the specific enthalpy
\begin{equation}
\frac{dh}{dR} (R )= \frac{w^2(R)-1}{R} \; ,
\end{equation}
with the usual meaning of $w$.

We now proceed as in the isothermal case, making use of the same dimensionless variables, with the difference that the scaling density is now defined simply as $\rho_1 = M_1/(4 \pi R_1^3)$.  In this way we come to the dimensionless system
\begin{equation}\label{polisystem} \begin{split}
\frac{d \lambda}{ds} (s) & = s^2 y(s)F(h(s)) g(\lambda(s)) \\
\frac{dh}{ds} (s) & = \frac{w^2(s)-1}{s} \; ,
\end{split} \end{equation}
where the functions $y$ and $w$ appearing on the right hand side are related to the unknowns $\lambda$ and $h$ by the invertible relations:
\begin{equation} \begin{split}
\lambda & =  sw \\
h & = \tilde{k}y^{\gamma -1} \; ,
\end{split} \end{equation}
where
\begin{equation}\label{ktildedef}
\tilde{k} \vcentcolon = \frac{\gamma}{\gamma-1}\frac{\rho_1^{\gamma-1}}{V_d^2} k
\end{equation}
is a dimensionless version of the constant $k$ in equation \eqref{polidef}.

A proper asymptotic analysis of equations \eqref{polisystem} is much more complex than in the isothermal case. We circumvent this difficulty pretending that there is a very small inner isothermal core, out of which equations can be integrated numerically straightforwardly. Because of equation \eqref{polidef} and recalling that $\gamma > 1$, a constant temperature in the core also implies a constant density and pressure there. Following what we already know from the isothermal case, this in turn implies that the core is in corotation with the disc ($w = 1$) and that the AMD is proportional to $\lambda^2$ there, formally in contrast with our hypotheses. However, as long as the enclosed mass of the core is small, this modification of the AMD can be regarded as irrelevant from the point of view of the mass and angular momentum budget (in significant contrast with our isothermal `wind' models) and we can therefore safely adopt this simplification.

Since this experiment is intended as an alternative to the wind model of Section \ref{sec::WindModels}, we use here the `primordial' AMD \eqref{truncatedexp}. We use a small core radius $R_c = 1 \; \textrm{kpc}$ and a core temperature $T_c  = 2.22 \times 10^6\; \textrm{K}$, while $k$ is chosen to match the outer boundary condition $\lambda(s_\textrm{max}) = \lambda_\textrm{max}$. To model the effect of small deviations from isothermality, we use $\gamma = 1.1$, close to the isothermal value $\gamma = 1$. On the other extreme, the isentropic value $\gamma = 5/3$ is far from isothermal and implies a much stronger temperature gradient: for this model, we had to fix a significantly larger core temperature $T_c = 6 \; T_\textrm{vir}$, to prevent the temperature from vanishing at a radius smaller than $s_\textrm{max}$. The equatorial temperature profiles for these two models are shown in Figure \ref{fig::Tplot}. Observational evidence in favour of a moderate temperature gradient, similar to that of our $\gamma = 1.1$ model, has recently been found, for the Milky Way, based on the comparison of oxygen emission lines in different ionization stages (\citealt{MillerBregman2015}). 

The kinematic properties of the two models are shown in Figure \ref{fig::Politropic}. The isentropic model ($\gamma = 5/3$) shows only little improvement with respect to the isothermal model of Figure \ref{fig::SuperVirial}. This is mainly because of the very large temperatures required in the inner regions in this case, which locally exacerbate the consequences of having a hot corona (cfr.\ Section \ref{sec::SuperVirialModels}). On the other hand, we find that the model with a mild temperature gradient ($\gamma = 1.1$) gives results that are, in the inner region, pretty similar to those of the extreme wind model (cfr.\ Figure \ref{fig::Wind}, lower panels): in particular, the specific angular momentum becomes equal to the disc average close to the edge of the disc. The (quite unexpected) conclusion is that very different strategies to modify our models (according to very different kinds of feedback) have a similar impact on the kinematics of a super-virial corona: in the most favourable conditions, they only bring it to marginal consistency with being a possible driver of the inside-out growth of the disc.

To facilitate comparisons, in Table \ref{table::summary} we give, for all the models considered so far, the radius $R_\textrm{cross}$ at which the specific angular momentum of the corona becomes larger than the average specific angular momentum of the disc (in Figures \ref{fig::Fig1}, \ref{fig::SuperVirial}, \ref{fig::Wind} and \ref{fig::Politropic}, the radius at which the horizontal dashed line is crossed). Lower values of $R_\textrm{cross}$ indicate that the disc is more likely to grow inside-out as a consequence of coronal accretion. The minimal models (with a tidal-torque-driven AMD and sub-virial temperatures) have $R_\textrm{cross} < 10 \; \textrm{kpc}$, irrespective of the details, while in all the other cases $R_\textrm{cross}$ is significantly larger.

\begin{table}
\centering
\caption{For each model in this paper, the radius $R_\textrm{cross}$ is given at which the specific angular momentum of the corona becomes larger than the average specific angular momentum of the disc (cfr. Figures \ref{fig::Fig1}, \ref{fig::SuperVirial}, \ref{fig::Wind} and \ref{fig::Politropic}). Ranges (min-max) quantify the dependence on boundary conditions. The smaller $R_\textrm{cross}$, the easier for the disc to accrete coronal gas with large specific angular momentum, which is needed to sustain inside-out growth.}\label{table::summary}
\begin{tabular}{cccc}
\hline
Model & Figure & \hspace{6.mm}$R_\textrm{cross}$ & \hspace{-6.mm}(kpc) \\
&& min & max \\
\hline
Minimal & \\
$a = 4$ & \ref{fig::Fig1} & 8.8 & 9.3\\
$a = 5$ & \ref{fig::Fig1} & 7.2 & 7.5\\
$a = 6$ & \ref{fig::Fig1} & 6.6 & 6.8\\
\hline
Supervirial & \\
No wind, isothermal & \ref{fig::SuperVirial} & 89 & 100 \\
Wind ($m = 0.4$)& \ref{fig::Wind} & 26 & 32\\
Wind ($m = 0.1$) & \ref{fig::Wind} & 15 & 16\\
Politropic ($\gamma = 1.1$) & \ref{fig::Politropic} & 16 & 17\\
Politropic ($\gamma = 5/3$) & \ref{fig::Politropic} & 67 & 73\\
\hline
\end{tabular}
\end{table}

Caution is obviously necessary in drawing conclusions. However, assuming that the observation of a super-virial temperature of the inner corona of the Milky Way is correct, the fact that it is so difficult to reconcile it with the expectations of an inside-out growing model may be suggestive that the radial growth of the Milky Way is, at the present epoch, approaching a halt. Even if confirmed, it is difficult to say whether this property is general or peculiar to our Galaxy. Caution in generalizations is particularly needed, especially considering the large observed scatter in the instantaneous radial growth rates of the discs of spiral galaxies, in a range of stellar masses comparable to that of our Galaxy (\citealt{Pezzulli+15}).

\section{The impact of the choice of the potential}\label{sec::Potential}
We finally assess here the impact of relaxing our approximation of a spherically symmetric logarithmic potential. As a test-bed for comparisons, we will focus on the case of a sub-virial isothermal corona with a primordial AMD (see Section \ref{sec::CosmoAMD}).

If the potential $\Phi$ is an arbitrary function of cylindrical coordinates $(R,z)$, equations \eqref{newsystem} generalize to
\begin{equation}\label{newsystemplus}
\begin{split}
\frac{dl}{dR}(R) & = 4 \pi \chi(R) R^2 \frac{\rho_0(R)}{\psi(l(R))} \\
\frac{d\rho_0}{dR}(R) & = \frac{1}{c_s^2} \frac{\rho_0(R)}{R} \left( \left( \frac{l(R)}{R} \right)^2 - V_\textrm{cen}^2(R) \right) \; ,
\end{split}
\end{equation}
where $c_s$ is the isothermal sound speed (as in Section \ref{sec::Basics}), while the dimensionless factor $\chi$ is no longer constant (as in equation \ref{defk}), but a function of radius
\begin{equation}\label{chidefplus}
\chi(R) \vcentcolon = \frac{1}{R}\int_0^{+\infty} \exp \left( - \frac{\Phi(R, z)-\Phi(R, 0)}{c_s^2} \right) dz \;
\end{equation}
and finally $V_\textrm{cen}$ is the velocity of \emph{centrifugal equilibrium} at the equator
\begin{equation}\label{Vcendef}
V_\textrm{cen}(R) \vcentcolon = \sqrt{R \frac{\partial \Phi}{\partial R} (R, 0)} \; ,
\end{equation}
which is a constant in the case of a logarithmic potential (equation \ref{logpot}), but a function of $R$ in more general cases.

Even in the absence of an intrinsic velocity scale, it is useful to put the problem in dimensionless form. To this purpose, an arbitrary \emph{reference} velocity $V_d$ must be chosen. In the following, we adopt $V_d = 240 \; \textrm{km} \; \textrm{s}^{-1}$, which facilitates comparisons with the rest of the paper. We then formally adopt the same positions \eqref{adef}, \eqref{scalings}, \eqref{dimensionlessvariables}, \eqref{gdef}, with the only difference that we remove the factor $\chi$ (which is now not constant) from the definition of $\rho_1$ (similarly to what we did in Section \ref{sec::PoliModels}). The system \eqref{newsystemplus} can then be written as
\begin{equation}\label{dimensionlessplus}
\begin{split}
\frac{d \lambda}{ds}(s) & = s^2 y \chi(R_1 s) g(\lambda(s)) \\
\frac{dy}{ds}(s) & = a \frac{y(s)}{s} \left( \left(\frac{\lambda(s)}{s}\right)^2 - \left(\frac{V_\textrm{cen}(R_1s)}{V_d}\right)^2 \right) \; ,
\end{split}
\end{equation}
which is similar to the system \eqref{dimensionless} but for the presence of two functions of one variable $\chi$ and $V_\textrm{cen}$, which can be tabulated, using equations \eqref{chidefplus} and \eqref{Vcendef}, for an adopted gravitational potential $\Phi$ and isothermal sound speed $c_s$.

\begin{figure*}
\centering
\includegraphics[width=18cm]{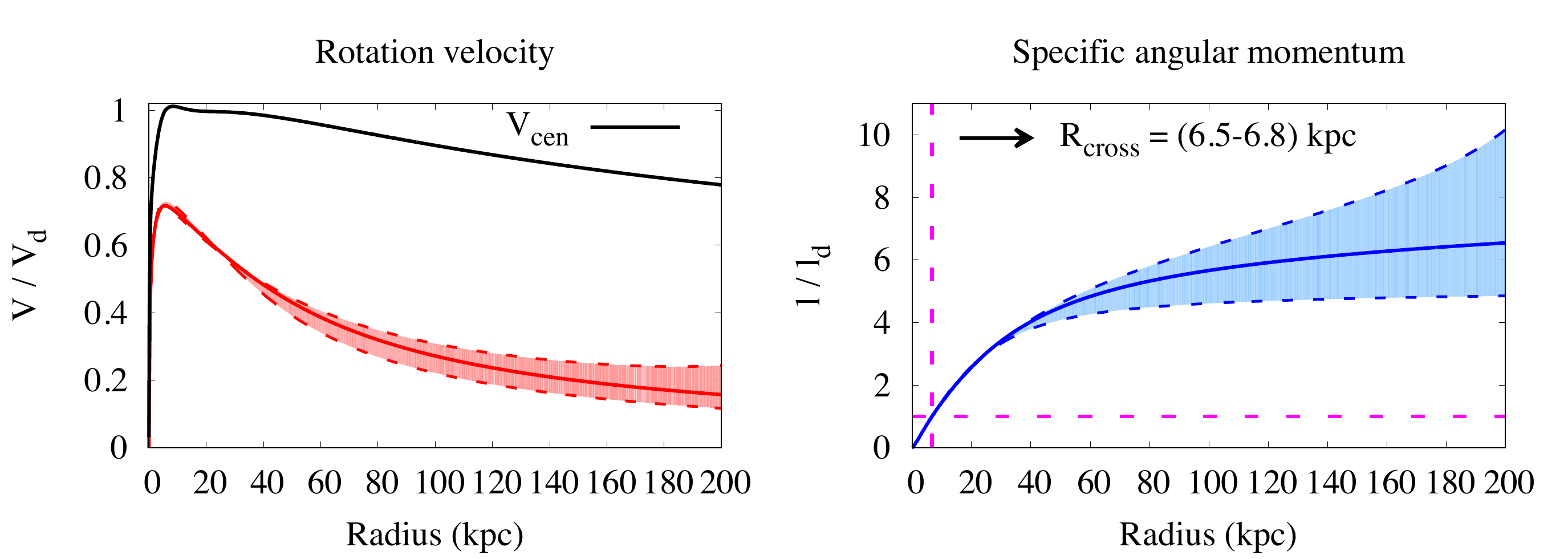}
\caption{Similar to the bottom panel of Figure \ref{fig::Fig1}, but adopting a more refined choice for the gravitational potential (see text). In the left-hand panel, rotation velocities are normalized to the \emph{reference} speed $V_d = 240 \; \textrm{km} \; \textrm{s}^{-1}$. The solid black line shows the speed of centrifugal equilibrium, which approximately equals $V_d$ only in the range $5 \lesssim R/\textrm{kpc} \lesssim 25$. In the innermost $\sim 5$ kpc, the rotation velocity of the corona steeply rises from zero to a maximum value, as the centrifugal speed does, because of the hydrodynamic bound $V < V_\textrm{cen}$. All other properties, including the value of $R_\textrm{cross}$, are remarkably similar to those estimated adopting a simplified logarithmic potential.\label{fig::Fig8}}
\end{figure*}

As anticipated in Section \ref{sec::inversion}, the equations are singular in the origin, irrespective of the potential, due to the centrifugal barrier $l^2/R^3$. Unfortunately, a general analytic treatment of the singularity is not possible for an arbitrary potential. To overcome this, we follow a strategy similar to that of Section \ref{sec::PoliModels}. We make the approximation that the equatorial density is constant within some very small radius $R_c$, so that $y = y_c$ for $0 < s < s_c \vcentcolon = R_c/R_1$. To be sensitive to the effect of a rapidly rising centrifugal speed $V_\textrm{cen}$, we adopt here a core radius $R_c = 100 \; \textrm{pc}$, 10 times smaller than the one used in Section \ref{sec::PoliModels}. The specific angular momentum at the core radius $\lambda_c = \lambda(s_c)$ is assigned by imposing that the mass enclosed within $0 < s < s_c$ coincides with the integral of the AMD in the range $0 < \lambda < \lambda_c$. \footnote{In more detail, we require \begin{displaymath} \frac{1}{M_1} \int_0^{l_1 \lambda_c} \psi(l) dl = y_c \int_0^{s_c} s^2 \chi(R_1s) ds \; ,\end{displaymath} where the integral in the right-hand side is evaluated after appropriately interpolating the function $\chi$ over the relevant radial range. Note that the optimal interpolation scheme can depend on the shape of the potential.} Then we numerically integrate equations \eqref{dimensionlessplus} from the core outwards, leaving $y_c$ as a free parameter, which is fixed a posteriori to match the desired boundary condition.

For our application, we consider a three-component gravitational potential generated by a disc, a bulge and a halo, following the method described by \cite{FB06}. We model the disc as a double exponential, with a scale-length $R_d = 2.5 \; \textrm{kpc}$, a scale-height $h_z = 300 \; \textrm{pc}$ and a mass $M_d = 4 \times 10^{10} \; \textrm{M}_\Sun$, while the bulge and the halo are modelled as flattened double-power-law distributions, with inner slope, outer slope, scale radius, maximum radius, aspect ratio and mass equal to $(-1\;,\; -5\;,\; 1.5 \; \textrm{kpc}\;,\; 4 \; \textrm{kpc}\;,\; 0.6\;,\; 10^{10} \; \textrm{M}_\Sun)$ for the bulge and $(-1\;,\; -3\;,\; 20 \; \textrm{kpc}\;,\; 200 \; \textrm{kpc}\;,\; 0.8\;,\; 1.5 \times 10^{12} \; \textrm{M}_\Sun)$ for the halo. Note that we do not pretend that the model above is the most accurate mass decomposition for the Milky Way, its purpose being limited to investigating the consequences of strong deviations from equation \eqref{logpot} with parameters that are plausible for our Galaxy. We finally notice that, because the rotation curve of the disc is not exactly constant, the approximation \eqref{expdiscldisc} for $l_d$ and therefore for $R_1$ can be improved. With our adopted potential, we compute $R_1 = 8 \; \textrm{kpc}$, slightly smaller than the fiducial value $R_1 = 8.3 \; \textrm{kpc} $ given in Section \ref{sec::dimensionless}, and we update all our scalings accordingly.

\subsection{Kinematics}

In Figure \ref{fig::Fig8} the kinematic properties of the resulting model are shown for the case $a = 6$ ($T = 0.5 \; T_\textrm{vir}$) and should be compared with the analogous calculations for a logarithmic potential (Figure \ref{fig::Fig1}, bottom panels). The black solid line in the left-hand panel of Figure \ref{fig::Fig8} is the radial profile of the centrifugal speed at the equator $V_\textrm{cen}$. We recall that this also coincides with the rotation velocity of a disc of cold gas in equilibrium in the potential $\Phi$. At variance with the logarithmic case (equation \ref{logpot}), where $V_\textrm{cen} = V_d$ at all radii by construction, here $V_\textrm{cen} = 0$ for $R = 0$. For $R > 0$, $V_\textrm{cen}$ shows a relatively steep rise, reaching $V_\textrm{cen} \simeq 240 \; \textrm{km} \; \textrm{s}^{-1}$ at $R \simeq 4 \; \textrm{kpc}$, remains approximately constant out to the observed extent of the disc and finally undergoes a significant decline towards the virial radius. These properties are consistent with current estimates of the rotation curve  of the Milky Way and similar external galaxies, as well as with theoretical expectations. The red line and shaded region show the model predictions for the rotation velocity of the corona. At variance with the predictions for the logarithmic potential (Figure \ref{fig::Fig1}, bottom-left panel), the rotation velocity of the corona here vanishes in the centre, as $V_\textrm{cen}$ does, it quickly rises with radius, reaching a maximum value and then it declines again, obeying $V < V_\textrm{cen}$ throughout. This behaviour is entirely explained by a basic theoretical consideration (see Section \ref{subsec::preliminary}): as long its radial pressure gradient is directed outwards, the hot gas is bound to rotate slower than the speed of centrifugal equilibrium; in particular, the fact that $V_\textrm{cen}$ vanishes in the origin requires $V$ to do so as well. However, outside an inner zone, approximately coincident with the spatial scale of the rise of $V_\textrm{cen}$, the coronal velocity reaches a value that is very close to that predicted under the assumption of a logarithmic potential and in most of the radial domain the two models are almost identical. The common behaviour at intermediate radii is not surprising, since $V_\textrm{cen}$ is constant -- or approximately constant -- there in both cases, while the similarity in the outer regions is to be ascribed to boundary conditions.

Key to the aim of this paper, the angular momentum profile is shown in the right-hand panel of Figure \ref{fig::Fig8}, to be compared with the bottom-right panel of Figure \ref{fig::Fig1}. As a consequence of the subtle differences in the rotation velocity profile, the specific angular momentum in the innermost regions is also slightly smaller than in the case of a logarithmic potential. At the same time (and for a very similar reason), the average specific angular momentum of the disc $l_d$, which the profile is normalized to, is also slightly smaller (see the estimate of $R_1$ above). Both effects are very small and, in addition, they tend to compensate each other, with the result that the predicted crossing radius $R_\textrm{cross}$ varies by less than 2 \% with respect to our previous simpler estimate.

The comparison above confirms that the details of the shape of the gravitational potential have a negligible impact in estimating the ability of galactic coronae to sustain the inside-out growth of discs and that the first-order approximation \eqref{logpot} is entirely sufficient to this end. More refined treatments, as the one described in this section, can however be relevant for fine comparisons of the predicted inner rotation velocity of the hot gas with future observations with a spatial resolution better than a few kpc.

\subsection{Density distribution}
\begin{figure}
\centering
\includegraphics[width=8.5cm]{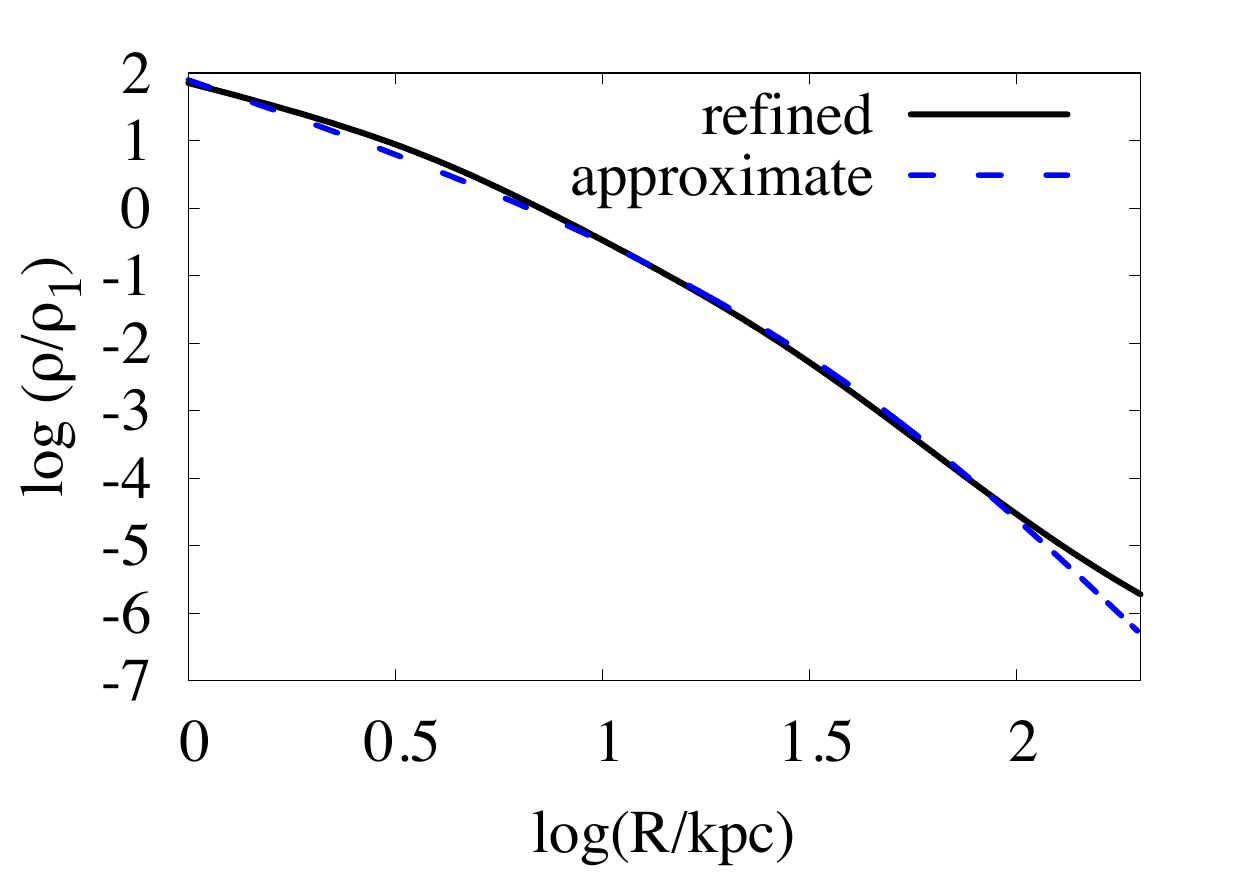}
\caption{Equatorial density profile for the \emph{refined} model of Figure \ref{fig::Fig8} (solid black line), compared with the \emph{approximate} predictions assuming a logarithmic potential (dashed blue line). The analytic model is an excellent approximation of the refined one, with minor discrepancies mostly at $R > 100 \; \textrm{kpc}$.}\label{fig::Fig9}
\end{figure}

\begin{figure}
\centering
\includegraphics[width=8.5cm]{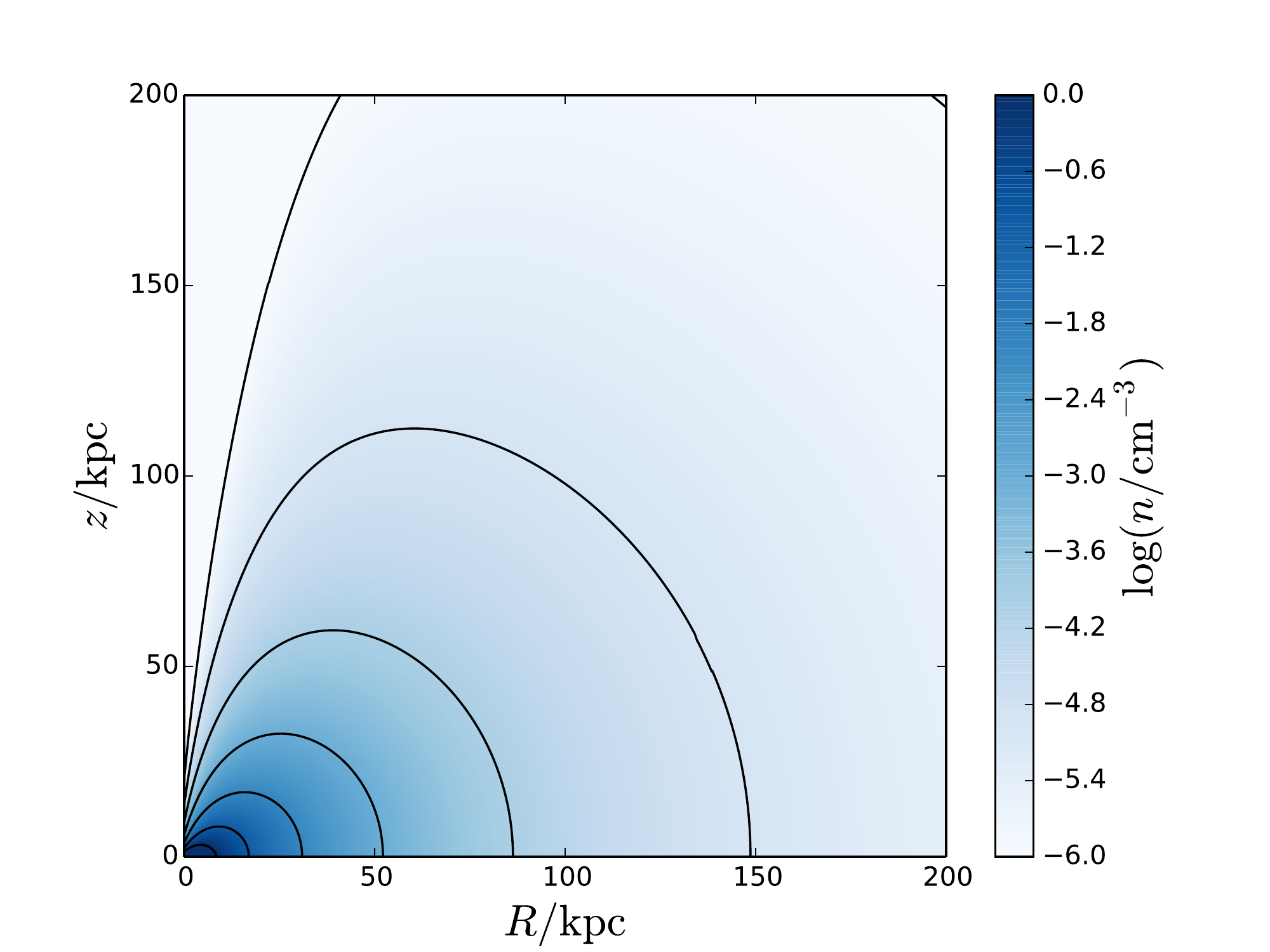}
\caption{Similar to Figure \ref{fig::IsoModelContours}, but assuming the refined gravitational potential described in Section \ref{sec::Potential}. Differences are small and limited to the outermost region. Larger effects may come from the removal of the barotropic assumption (see Section \ref{sec::DiscussionBarotropic}).}\label{fig::Fig10}
\end{figure}

For completeness, we briefly discuss here how the improved potential affects predictions of the density distribution of the corona. Figure \ref{fig::Fig9} compares the equatorial density profiles predicted, for $a = 6$, with different assumptions for the potential. Note that, for consistency, we adopt here a common definition $\rho_1 = M_1/4 \pi R_1^3$ in both cases. The two profiles are remarkably similar to each other. Notably, the analytic prediction for the inner slope of the density profile, which is rigorously valid only in the limit $R \to 0$ and for a logarithmic potential, is very accurate also in the non-logarithmic case. This suggests that the asymptotic expansions presented in Section \ref{sec::inversion} provide useful quick estimates of the expected coronal structure even in the absence of a detailed knowledge of the potential. On the other hand, in the outer regions ($R > 100 \; \textrm{kpc}$), the density profile with the refined potential is slightly shallower than what predicted using a logarithmic approximation. Since the rotation velocity at these radii is much smaller than the centrifugal speed, the effect can be easily understood in terms of hydrostatic equilibrium: a less intense outer gravitational field requires smaller pressure gradients, which, at fixed temperature, implies a shallower density profile. The same effect can also be seen in Figure \ref{fig::Fig10}, which is very similar to the analogous Figure \ref{fig::IsoModelContours} but for the outermost density contour.

\section{Discussion}\label{sec::Discussion}
\subsection{Deviations from barotropic equilibrium}\label{sec::DiscussionBarotropic}

A limitation of our approach comes from the assumption that the equilibrium is barotropic (pressure and the density are stratified on the same surfaces). As already recalled in Section \ref{sec::starteq}, all barotropic equilibria are forced to rotate on cylinders. Therefore, at any given cylindrical radius $R$, there will be a critical height $z_c(R)$ above which the rotation becomes super-centrifugal, implying a radially increasing pressure (and density). This is the main reason why our models can have very low densities (and large density gradients) at small $R$ and large $|z|$ (see Figures \ref{fig::IsoModelContours} and \ref{fig::Fig10}). More general baroclinic models can have a more regular behaviour near the rotation axis (e.g.\ \citealt{Barnabe+06}), implying that our predictions in this region should be treated with caution. We plan to address this aspect in more detail in future work.

Real cosmological coronae will also depart, at some level, from equilibrium (either barotropic or baroclinic). Such deviations are dynamically important only if they involve very large velocities, comparable to the virial speed $V_d$. Non-equilibrium motions of this magnitude are indeed expected when the corona is formed, though the associated energy should be dissipated efficiently, especially in the inner regions where the dynamical time is shorter, and have a small impact on the final configuration of the hot gas (e.g.\ \citealt{Voit+03}). Primordial turbulence could also, in principle, constitute a source of angular momentum mixing, therefore altering the AMD of the hot gas. Early studies (e.g.\ \citealt{Mestel1963}) considered this possibility and found that the angular momentum redistribution due to \emph{turbulent viscosity} should be negligible. However, classical arguments do not apply if significant levels of turbulence are constantly maintained by continuous injection of mechanical energy by feedback processes. Galactic fountains are expected to mix cold gas from the disc with the hot gas in a low $|z|$ layer of the corona, with consequences on the angular momentum of both phases (e.g.\ \citealt{Marinacci+11}). As long as the spatial extent of the fountain cycle is moderate (1 to 10 kpc scale), as suggested by observations of the extraplanar gas in nearby galaxies and in the Milky Way (e.g.\ \citealt{Fraternali2002}; \citealt{Oosterloo07Lag}; \citealt{Marasco+11}), then the process mainly puts the disc in contact with its immediate surroundings, therefore leaving all our conclusions unchanged. Flows with much larger spatial extent are however sometimes invoked by cosmological simulations with very strong feedback (e.g.\ \citealt{Brook+12}) and may be worth more theoretical as well as observational investigation (see also Section \ref{sec::Evolution} below).

\subsection{Other sources of accretion}
Implicit to our investigation is the assumption that the majority of the gas accretion on to massive star forming galaxies comes from the slow condensation of the hot coronae which surround them. This is expected on a variety of theoretical grounds (e.g.\ \citealt{BirnboimDekel2003}; \citealt{Binney2004}; \citealt{DB06}; \citealt{Nelson+13}), while direct accretion of cold ($T \lesssim 10^4 \; \textrm{K}$) gas from the intergalactic medium is expected to be common mostly for low-mass galaxies and during the earliest stages of galaxy formation.

Observations help to constrain the amount of residual direct accretion of cold gas on star forming galaxies in the Local Universe (e.g.\ \citealt{Sanchez-AlmeidaReview} and references therein). Some supply of cold gas is occasionally brought to galaxy discs by minor mergers (e.g.\ \citealt{Sancisi+08}), but the average gas accretion rate from this channel is much smaller than what is needed to sustain the observed levels of star formation in the Local Universe (\citealt{DiTeodoro2014}).

Clouds of cold gas, kinematically consistent with being infalling on to the disc (the so-called \emph{high-velocity clouds}, or HVCs), exist around the Milky Way (\citealt{Oort1970}) as well as around external galaxies (e.g.\ \citealt{vdHSancisi88}), but they are invariably found only a few kpc away from the disc itself (e.g.\ \citealt{vanWoerdenWakker04}; \citealt{Fraternali2009}), which is hard to explain if they directly come from the intergalactic medium and rather suggests that they are associated with the condensation of the lower corona (\citealt{Fraternali+15}). HVCs are not massive and numerous enough to contribute significant gas accretion (e.g.\ \citealt{Putman2012} and references therein, though see \citealt{LehnerHowk11}). Larger accretion rates are associated with intermediate-velocity clouds (IVCs), which are even more closely connected to the disc and kinematically consistent with being condensed out of coronal gas (\citealt{Marasco+12}; \citealt{Marasco+13}).

It is finally possible that a large amount of infalling cold gas is almost invisible, either because it has a very low column density or because it is almost completely ionized. Low column density cold ionized gas in the haloes of nearby galaxies has been recently detected in absorption along sight-lines of background quasars (e.g.\ \citealt{Lehner+13}; \citealt{Churchill+13}; \citealt{Tumlinson+13}). The total mass of this phase is uncertain, estimates being sensitive to the details of photo-ionization modelling (e.g.\  \citealt{Werk+14}; \citealt{Stern+16}). Moreover, projection effects prevent to assess whether and how much of this gas is actually falling towards the central galaxy. The volume density is also very uncertain; we notice, however, that the models which predict the largest mass also require very low volume densities (two orders of magnitude below pressure equilibrium, see the discussion in \citealt{Werk+14}), which would make difficult for hypothetically infalling clouds to reach the disc without being ablated away by interactions with the hot medium (e.g.\ \citealt{Binney2004}).

\subsection{What long-term consequences to galaxy evolution?}\label{sec::Evolution}
As applications of our reconstruction method (Section \ref{sec::inversion}), we have presented in some detail a few examples (Sections \ref{sec::CosmoAMD} to \ref{sec::PoliModels}), chosen to encompass a range of very different possible situations. For each situation, we provided an essentially qualitative assessment: some coronal models are in principle able to sustain the inside-out growth of an embedded disc, while some others are not.

A fully quantitative account of the role of coronal accretion in feeding the inside-out growth of discs also requires the knowledge of \emph{how much} mass and angular momentum accretes, per unit time, from the corona on to the disc. Furthermore, it would require a \emph{continuum} of models and a prescription to specify whether and how they are connected in an evolutionary sequence. These goals cannot be achieved without knowing the \emph{accretion rate} (how much coronal gas actually cools down on to the disc per unit time) and the \emph{accretion profile} (the radial distribution of the accretion rate). The accretion rate and profile are determined by the local efficiency with which coronal gas condenses on to the disc. This depends on several intertwined factors, including density, temperature and metallicity (e.g.\ \citealt{SutherlandDopita1993}), heating sources of any kind, photo-ionization from both the UV background and local sources (e.g.\ \citealt{Efstathiou1992}; \citealt{Wiersma+09}; \citealt{Cantalupo2010}) and the complex interaction of cold and hot gas at the disc/corona interface (e.g.\ \citealt{Fraternali2016Review} and references therein), which is sensitive to many poorly known physical processes (e.g.\ turbulence, thermal conduction and magnetic fields) and acts at spatial scales too small to be resolved by current cosmological simulations. Furthermore, as we have seen, predictions can depend on spatially-resolved properties of thermal and mechanical feedback, which are also far away from being theoretically or observationally well settled. An adequate understanding of all these factors and of their mutual interplay is a major challenge of modern astrophysics and is outside the scope of this work.

Any semi-analytical model of galaxy evolution, including prescriptions for one or more of the processes mentioned above, could take advantage of the reconstruction techniques described in this paper to follow the detailed evolution of the corona self-consistently. As long as the processes involved do not vary dramatically on time-scales shorter than the sound-crossing time of the inner corona (typically, $\sim 100 \; \textrm{Myr}$), then the latter would be found, at any time, in a quasi-equilibrium state, which, thanks to our formalism, can be computed, at each time-step, in a very efficient way \footnote{Each computation just requires the integration of two equations and one optimization loop; see Section \ref{sec::inversion} for details.}. Very short time-scale, out-of-equilibrium events (such as AGN-driven injections of thermal and/or mechanical energy) cannot be followed in their detailed development and should be treated as impulsive discontinuities in the evolution of the AMD and/or thermal structure of the corona. The density and rotation velocity profiles of the corona, thus found, can then be used as an input to derive, for instance, the accretion rate and profile at the following time step according to the chosen prescription. We also notice that this approach potentially gives, as a by-product, predictions for the chemical evolution of the disc, because of the tight relationship between abundance gradients and the kinematics of the accreting gas (e.g.\ \citealt{PF16} and references therein).

Without entering into the details of any specific scenario, we can use the examples provided in this work to outline what general expectations for an evolutionary model should be.  As long as a corona is kept sub-virial (that is, as long as negligible energy other than gravitational is injected into it, or if cooling is very efficient), it should be relatively easy for coronal gas to sustain the inside-out growth of the disc, provided that the accretion profile is such that enough material from relatively outer regions is harvested (e.g.\ \citealt{Marasco+12}; \citealt{Bouche+13Sci}). Both accretion itself and mechanical feedback are expected to remove low-$l$ material from the inner corona, further facilitating the process. Note that, to be effective in this sense, mechanical feedback is not required to completely expel the low-$l$ corona from the halo: `recycled winds' (or large-scale fountains) would induce the mixing of material with different specific angular momentum (e.g.\ \citealt{Brook+12}), altering the \emph{shape} of the AMD in the sense required to facilitate inside-out growth (this could be modelled, for instance, as an increase of $p$, in equation \ref{SS05}, at fixed $l_1$). Thermal feedback, on the other hand, would induce the expansion of the corona and the migration of high-$l$ gas towards far, unaccessible regions, resulting either into a halt to the accretion or the shrinkage of the disc. The latter effect could be prevented only if thermal energy deposition is centrally concentrated enough, or if vigorous mechanical feedback is acting in concert, or a combination of the two.

\section{Summary and conclusions}\label{sec::Summary}

The discs of spiral galaxies are believed to be continuously accreting mass and angular momentum from the large cosmological coronae of hot and diffuse gas that surround them. Cosmological coronae are indeed expected to have a specific angular momentum slightly larger than that of the embedded galaxies, which makes them \emph{potential} drivers of the angular momentum growth (or inside-out growth) of galaxy discs. However, since coronae are so extended, it is not obvious that their angular momentum is spatially concentrated enough to be actually available for accretion.

In this paper, we investigated the link between the \emph{spatial} distribution of the angular momentum of galactic coronae (which determines whether a galaxy accreting from this reservoir can grow inside-out) and the way angular momentum is distributed \emph{by mass} (the AMD, which is the quantity that is most readily predicted by cosmological models of structure formation). We developed an analytical method to reconstruct the kinematics and the density distribution of a hot gaseous halo in barotropic rotating equilibrium within a galactic gravitational potential from the knowledge of its AMD. Our method is largely insensitive to the detailed shape of the potential, which makes it easy to implement in semi-analytical models of galaxy evolution and allows robust predictions to be made even in the absence of detailed mass decompositions.

We have found that a hot rotating corona is a viable driver for the inside-out growth of a galactic disc, provided that it has an approximately exponential AMD (as predicted by tidal torque theory) and a slightly sub-virial temperature (as expected from the virial theorem for a rotating gas in isolation). In this \emph{minimal model}, the rotation velocity of the corona is slightly sub-centrifugal (smaller than, but close to, the rotation velocity of the disc) in the inner (galaxy-scale) regions and it declines substantially towards the virial radius. The prediction for the inner rotation velocity agrees with expectations from chemical evolution models and also with recent observational estimates of the kinematics of the corona of the Milky Way.

We then explored how our predictions change due to modifications of our minimal assumptions. Raising the coronal temperature to a super-virial value moves the high-angular-momentum gas, needed to sustain inside-out growth, far away from the disc, where it is unavailable for direct accretion. Conversely, the removal of low-angular-momentum gas from the corona, for instance by selective mechanical feedback, facilitates accretion-driven inside-out growth and can marginally compensate, in favourable conditions, the effect of a super-virial temperature. The predictions of this scenario are however degenerate with those of purely thermal feedback with a moderate temperature gradient.

If applied to the Milky Way, our findings suggest that our Galaxy may be, at the present epoch, close to the end of inside-out growth, which is not inconsistent with the relatively low \emph{instantaneous} radial growth rates measured in nearby, relatively massive, spiral galaxies.

Future comparisons with observations of the coronae of external spiral galaxies will help to constrain how easy or difficult is for galactic coronae to feed the angular momentum growth of discs. Comparing our models with the easier-to-measure coronae of ellipticals would also be interesting, to assess whether and how the latter have been shaped by the feedback mechanisms that are thought to be responsible for the halt of their star formation.

\section*{Acknowledgements}
GP is grateful to Aura Obreja for directing his attention to the work of \cite{SharmaSteinmetz2005}, to Rodrigo Ibata for carefully reading an earlier version of this work and to Payel Das for useful suggestions on the presentation. The authors also thank the referee for useful comments and in particular for suggesting the test of Section \ref{sec::Potential}. GP and FF acknowledge financial support from PRIN MIUR 2010-2011, project `The Chemical and Dynamical Evolution of the Milky Way and Local Group Galaxies', prot. 2010LY5N2T. This work was partly funded by the Marco Polo Programme (University of Bologna), by the European Research Council under the European Union's Seventh Framework Programme (FP7/2007-2013)/ERC grant agreement no.\ 321067 and by the Swiss National Foundation grant PP00P2\_163824.

\bibliographystyle{mn2e}
\bibliography{mybib}{}

\label{lastpage}
\end{document}